\documentclass[12pt,a4paper]{article}
\usepackage{amssymb}
\usepackage{graphicx}

\usepackage[russian]{babel}

\makeatletter
\newcount\@hour\newcount\@minute\chardef\@x10\chardef\@xv60
\def\tcitime{
\def\@time{%
  \@minute\time\@hour\@minute\divide\@hour\@xv
  \ifnum\@hour<\@x 0\fi\the\@hour:%
  \multiply\@hour\@xv\advance\@minute-\@hour
  \ifnum\@minute<\@x 0\fi\the\@minute
  }}%
\def\QCTOpt[#1]#2{%
  \def\QCTOptB{#1}
  \def\QCTOptA{#2}
}
\def\QCTNOpt#1{%
  \def\QCTOptA{#1}
  \let\QCTOptB\empty
}
\def\Qct{%
  \@ifnextchar[{%
    \QCTOpt}{\QCTNOpt}
}
\def\QCBOpt[#1]#2{%
  \def\QCBOptB{#1}
  \def\QCBOptA{#2}
}
\def\QCBNOpt#1{%
  \def\QCBOptA{#1}
  \let\QCBOptB\empty
}
\def\Qcb{%
  \@ifnextchar[{%
    \QCBOpt}{\QCBNOpt}
}
\def\PrepCapArgs{%
  \ifx\QCBOptA\empty
    \ifx\QCTOptA\empty
      {}%
    \else
      \ifx\QCTOptB\empty
        {\QCTOptA}%
      \else
        [\QCTOptB]{\QCTOptA}%
      \fi
    \fi
  \else
    \ifx\QCBOptA\empty
      {}%
    \else
      \ifx\QCBOptB\empty
        {\QCBOptA}%
      \else
        [\QCBOptB]{\QCBOptA}%
      \fi
    \fi
  \fi
}
\newcount\GRAPHICSTYPE
\GRAPHICSTYPE=\z@
\def\GRAPHICSPS#1{%
 \ifcase\GRAPHICSTYPE
   \special{ps: #1}%
 \or
   \special{language "PS", include "#1"}%
 \fi
}%
\def\graffile#1#2#3#4{%
    \leavevmode
    \raise -#4 \BOXTHEFRAME{%
        \hbox to #2{\raise #3\hbox{\null #1}}}%
}%
\def\draftbox#1#2#3#4{%
 \leavevmode\raise -#4 \hbox{%
  \frame{\rlap{\protect\tiny #1}\hbox to #2%
   {\vrule height#3 width\z@ depth\z@\hfil}%
  }%
 }%
}%
\newcount\draft
\draft=\z@

\newif\ifwasdraft
\wasdraftfalse

\def\GRAPHIC#1#2#3#4#5{%
 \ifnum\draft=\@ne\draftbox{#2}{#3}{#4}{#5}%
  \else\graffile{#1}{#3}{#4}{#5}%
  \fi
 }%
\def\addtoLaTeXparams#1{%
    \edef\LaTeXparams{\LaTeXparams #1}}%
\newif\ifBoxFrame \BoxFramefalse
\newif\ifOverFrame \OverFramefalse
\newif\ifUnderFrame \UnderFramefalse

\def\BOXTHEFRAME#1{%
   \hbox{%
      \ifBoxFrame
         \frame{#1}%
      \else
         {#1}%
      \fi
   }%
}

\def\doFRAMEparams#1{\BoxFramefalse\OverFramefalse\UnderFramefalse\readFRAMEparams#1\end}%
\def\readFRAMEparams#1{%
 \ifx#1\end%
  \let\next=\relax
  \else
  \ifx#1i\dispkind=\z@\fi
  \ifx#1d\dispkind=\@ne\fi
  \ifx#1f\dispkind=\tw@\fi
  \ifx#1t\addtoLaTeXparams{t}\fi
  \ifx#1b\addtoLaTeXparams{b}\fi
  \ifx#1p\addtoLaTeXparams{p}\fi
  \ifx#1h\addtoLaTeXparams{h}\fi
  \ifx#1X\BoxFrametrue\fi
  \ifx#1O\OverFrametrue\fi
  \ifx#1U\UnderFrametrue\fi
  \ifx#1w
    \ifnum\draft=1\wasdrafttrue\else\wasdraftfalse\fi
    \draft=\@ne
  \fi
  \let\next=\readFRAMEparams
  \fi
 \next
 }%
\def\IFRAME#1#2#3#4#5#6{%
      \bgroup
      \let\QCTOptA\empty
      \let\QCTOptB\empty
      \let\QCBOptA\empty
      \let\QCBOptB\empty
      #6%
      \parindent=0pt%
      \leftskip=0pt
      \rightskip=0pt
      \setbox0 = \hbox{\QCBOptA}%
      \@tempdima = #1\relax
      \ifOverFrame
          \typeout{This is not implemented yet}%
          \show\HELP
      \else
         \ifdim\wd0>\@tempdima
            \advance\@tempdima by \@tempdima
            \ifdim\wd0 >\@tempdima
               \textwidth=\@tempdima
               \setbox1 =\vbox{%
                  \noindent\hbox to \@tempdima{\hfill\GRAPHIC{#5}{#4}{#1}{#2}{#3}\hfill}\\%
                  \noindent\hbox to \@tempdima{\parbox[b]{\@tempdima}{\QCBOptA}}%
               }%
               \wd1=\@tempdima
            \else
               \textwidth=\wd0
               \setbox1 =\vbox{%
                 \noindent\hbox to \wd0{\hfill\GRAPHIC{#5}{#4}{#1}{#2}{#3}\hfill}\\%
                 \noindent\hbox{\QCBOptA}%
               }%
               \wd1=\wd0
            \fi
         \else
            \ifdim\wd0>0pt
              \hsize=\@tempdima
              \setbox1 =\vbox{%
                \unskip\GRAPHIC{#5}{#4}{#1}{#2}{0pt}%
                \break
                \unskip\hbox to \@tempdima{\hfill \QCBOptA\hfill}%
              }%
              \wd1=\@tempdima
           \else
              \hsize=\@tempdima
              \setbox1 =\vbox{%
                \unskip\GRAPHIC{#5}{#4}{#1}{#2}{0pt}%
              }%
              \wd1=\@tempdima
           \fi
         \fi
         \@tempdimb=\ht1
         \advance\@tempdimb by \dp1
         \advance\@tempdimb by -#2%
         \advance\@tempdimb by #3%
         \leavevmode
         \raise -\@tempdimb \hbox{\box1}%
      \fi
      \egroup%
}%
\def\DFRAME#1#2#3#4#5{%
 \begin{center}
     \let\QCTOptA\empty
     \let\QCTOptB\empty
     \let\QCBOptA\empty
     \let\QCBOptB\empty
     \ifOverFrame 
        #5\QCTOptA\par
     \fi
     \GRAPHIC{#4}{#3}{#1}{#2}{\z@}
     \ifUnderFrame 
        \par #5\QCBOptA
     \fi
 \end{center}%
 }%
\def\FFRAME#1#2#3#4#5#6#7{%
 \begin{figure}[#1]%
  \let\QCTOptA\empty
  \let\QCTOptB\empty
  \let\QCBOptA\empty
  \let\QCBOptB\empty
  \ifOverFrame
    #4
    \ifx\QCTOptA\empty
    \else
      \ifx\QCTOptB\empty
        \caption{\QCTOptA}%
      \else
        \caption[\QCTOptB]{\QCTOptA}%
      \fi
    \fi
    \ifUnderFrame\else
      \label{#5}%
    \fi
  \else
    \UnderFrametrue%
  \fi
  \begin{center}\GRAPHIC{#7}{#6}{#2}{#3}{\z@}\end{center}%
  \ifUnderFrame
    #4
    \ifx\QCBOptA\empty
      \caption{}%
    \else
      \ifx\QCBOptB\empty
        \caption{\QCBOptA}%
      \else
        \caption[\QCBOptB]{\QCBOptA}%
      \fi
    \fi
    \label{#5}%
  \fi
  \end{figure}%
 }%
\newcount\dispkind%
\def\FRAME#1#2#3#4#5#6#7#8{%
 \ifnum\draft=\@ne
   \wasdrafttrue
 \else
   \wasdraftfalse%
 \fi
 \def\LaTeXparams{}%
 \dispkind=\z@
 \def\LaTeXparams{}%
 \doFRAMEparams{#1}%
 \ifnum\dispkind=\z@\IFRAME{#2}{#3}{#4}{#7}{#8}{#5}\else
  \ifnum\dispkind=\@ne\DFRAME{#2}{#3}{#7}{#8}{#5}\else
   \ifnum\dispkind=\tw@
    \edef\@tempa{\noexpand\FFRAME{\LaTeXparams}}%
    \@tempa{#2}{#3}{#5}{#6}{#7}{#8}%
    \fi
   \fi
  \fi
  \ifwasdraft\draft=1\else\draft=0\fi{}%
 }%
\def\TEXUX#1{"texux"}

\def\func#1{\mathop{\rm #1}}%
\def\limfunc#1{\mathop{\rm #1}}%

\long\def\QQQ#1#2{%
     \long\expandafter\def\csname#1\endcsname{#2}}%
\@ifundefined{QTP}{\def\QTP#1{}}{}
\@ifundefined{Qlb}{}{}
\@ifundefined{Qlt}{}{}
\long\def\QQA#1#2{}%
\def\QTR#1#2{{\csname#1\endcsname #2}}
\def\EXPAND#1[#2]#3{}%
\def\NOEXPAND#1[#2]#3{}%
\def\LaTeXparent#1{}%
\def\ChildStyles#1{}%
\def\ChildDefaults#1{}%
\def\QTagDef#1#2#3{}%
\def\QQfnmark#1{\footnotemark}

\@ifundefined{INDEX}{\def\INDEX#1#2{}{}}{}%
\@ifundefined{SUBINDEX}{\def\SUBINDEX#1#2#3{}{}{}}{}%
\def\initial#1{\bigbreak{\raggedright\large\bf #1}\kern 2\p@
   \penalty3000}%
\@ifundefined{ZZZ}{}{\makeatletter\input gnuindex.sty\makeatother\makeindex\makeatletter}%
\@ifundefined{abstract}{%
 \def\abstract{%
  \if@twocolumn
   \section*{Abstract (Not appropriate in this style!)}%
   \else \small 
   \begin{center}{\bf Abstract\vspace{-.5em}\vspace{\z@}}\end{center}%
   \quotation 
   \fi
  }%
 }{%
 }%
\@ifundefined{endabstract}{\def\endabstract
  {\if@twocolumn\else\endquotation\fi}}{}%
\@ifundefined{maketitle}{\def\maketitle#1{}}{}%
\@ifundefined{affiliation}{\def\affiliation#1{}}{}%
\@ifundefined{proof}{}{}%
\@ifundefined{endproof}{}{}%
\@ifundefined{newfield}{\def\newfield#1#2{}}{}%
\@ifundefined{chapter}{\def\chapter#1{\par(Chapter head:)#1\par }%
 \newcount\c@chapter}{}%
\@ifundefined{part}{\def\part#1{\par(Part head:)#1\par }}{}%
\@ifundefined{section}{\def\section#1{\par(Section head:)#1\par }}{}%
\@ifundefined{subsection}{\def\subsection#1%
 {\par(Subsection head:)#1\par }}{}%
\@ifundefined{subsubsection}{\def\subsubsection#1%
 {\par(Subsubsection head:)#1\par }}{}%
\@ifundefined{paragraph}{\def\paragraph#1%
 {\par(Subsubsubsection head:)#1\par }}{}%
\@ifundefined{subparagraph}{\def\subparagraph#1%
 {\par(Subsubsubsubsection head:)#1\par }}{}%
\@ifundefined{therefore}{}{}%
\@ifundefined{backepsilon}{}{}%
\@ifundefined{yen}{}{}%
\@ifundefined{registered}{%
   \def\registered{\relax\ifmmode{}\r@gistered
                    \else$\m@th\r@gistered$\fi}%
 \def\r@gistered{^{\ooalign
  {\hfil\raise.07ex\hbox{$\scriptstyle\rm\text{R}$}\hfil\crcr
  \mathhexbox20D}}}}{}%
\@ifundefined{Eth}{}{}%
\@ifundefined{eth}{}{}%
\@ifundefined{Thorn}{}{}%
\@ifundefined{thorn}{}{}%
\@ifundefined{degree}{}{}%
\newdimen\theight
\def\Column{%
 \vadjust{\setbox\z@=\hbox{\scriptsize\quad\quad tcol}%
  \theight=\ht\z@\advance\theight by \dp\z@\advance\theight by \lineskip
  \kern -\theight \vbox to \theight{%
   \rightline{\rlap{\box\z@}}%
   \vss
   }%
  }%
 }%
\def\qed{%
 \ifhmode\unskip\nobreak\fi\ifmmode\ifinner\else\hskip5\p@\fi\fi
 \hbox{\hskip5\p@\vrule width4\p@ height6\p@ depth1.5\p@\hskip\p@}%
 }%
\def\miss{\hbox{\vrule height2\p@ width 2\p@ depth\z@}}%
\def\tcol#1{{\baselineskip=6\p@ \vcenter{#1}} \Column}  %
\def\newfmtname{LaTeX2e}
\def\chkcompat{%
   \if@compatibility
   \else
     \usepackage{latexsym}
   \fi
}

\ifx\fmtname\newfmtname
  \DeclareOldFontCommand{\rm}{\normalfont\rmfamily}{\mathrm}
  \DeclareOldFontCommand{\sf}{\normalfont\sffamily}{\mathsf}
  \DeclareOldFontCommand{\tt}{\normalfont\ttfamily}{\mathtt}
  \DeclareOldFontCommand{\bf}{\normalfont\bfseries}{\mathbf}
  \DeclareOldFontCommand{\it}{\normalfont\itshape}{\mathit}
  \DeclareOldFontCommand{\sl}{\normalfont\slshape}{\@nomath\sl}
  \DeclareOldFontCommand{\sc}{\normalfont\scshape}{\@nomath\sc}
  \chkcompat
\fi

\@ifundefined{theorem}{}{}
\@ifundefined{lemma}{}{}
\@ifundefined{corollary}{}{}
\@ifundefined{conjecture}{}{}
\@ifundefined{proposition}{}{}
\@ifundefined{axiom}{}{}
\@ifundefined{remark}{}{}
\@ifundefined{example}{}{}
\@ifundefined{exercise}{}{}
\@ifundefined{definition}{}{}

\@ifundefined{mathletters}{%
  \newcounter{equationnumber}  
  \def\mathletters{%
     \addtocounter{equation}{1}
     \edef\@currentlabel{\theequation}%
     \setcounter{equationnumber}{\c@equation}
     \setcounter{equation}{0}%
     \edef\theequation{\@currentlabel\noexpand\alph{equation}}%
  }
  
}{}

\@ifundefined{BibTeX}{%
    \def\BibTeX{{\rm B\kern-.05em{\sc i\kern-.025em b}\kern-.08em
                 T\kern-.1667em\lower.7ex\hbox{E}\kern-.125emX}}}{}%
\@ifundefined{AmS}%
    {\def\AmS{{\protect\usefont{OMS}{cmsy}{m}{n}%
                A\kern-.1667em\lower.5ex\hbox{M}\kern-.125emS}}}{}%
\@ifundefined{AmSTeX}{}{}%
\let\DOTSI\relax
\def\RIfM@{\relax\ifmmode}%
\def\FN@{\futurelet\next}%
\newcount\intno@
\def\iint{\DOTSI\intno@\tw@\FN@\ints@}%
\def\iiint{\DOTSI\intno@\thr@@\FN@\ints@}%
\def\iiiint{\DOTSI\intno@4 \FN@\ints@}%
\def\idotsint{\DOTSI\intno@\z@\FN@\ints@}%
\def\ints@{\findlimits@\ints@@}%
\newif\iflimtoken@
\newif\iflimits@
\def\findlimits@{\limtoken@true\ifx\next\limits\limits@true
 \else\ifx\next\nolimits\limits@false\else
 \limtoken@false\ifx\ilimits@\nolimits\limits@false\else
 \ifinner\limits@false\else\limits@true\fi\fi\fi\fi}%
\def\multint@{\int\ifnum\intno@=\z@\intdots@                          
 \else\intkern@\fi                                                    
 \ifnum\intno@>\tw@\int\intkern@\fi                                   
 \ifnum\intno@>\thr@@\int\intkern@\fi                                 
 \int}
\def\multintlimits@{\intop\ifnum\intno@=\z@\intdots@\else\intkern@\fi
 \ifnum\intno@>\tw@\intop\intkern@\fi
 \ifnum\intno@>\thr@@\intop\intkern@\fi\intop}%
\def\intic@{%
    \mathchoice{\hskip.5em}{\hskip.4em}{\hskip.4em}{\hskip.4em}}%
\def\negintic@{\mathchoice
 {\hskip-.5em}{\hskip-.4em}{\hskip-.4em}{\hskip-.4em}}%
\def\ints@@{\iflimtoken@                                              
 \def\ints@@@{\iflimits@\negintic@
   \mathop{\intic@\multintlimits@}\limits                             
  \else\multint@\nolimits\fi                                          
  \eat@}
 \else                                                                
 \def\ints@@@{\iflimits@\negintic@
  \mathop{\intic@\multintlimits@}\limits\else
  \multint@\nolimits\fi}\fi\ints@@@}%
\def\intkern@{\mathchoice{\!\!\!}{\!\!}{\!\!}{\!\!}}%
\def\plaincdots@{\mathinner{\cdotp\cdotp\cdotp}}%
\def\intdots@{\mathchoice{\plaincdots@}%
 {{\cdotp}\mkern1.5mu{\cdotp}\mkern1.5mu{\cdotp}}%
 {{\cdotp}\mkern1mu{\cdotp}\mkern1mu{\cdotp}}%
 {{\cdotp}\mkern1mu{\cdotp}\mkern1mu{\cdotp}}}%
\def\RIfM@{\relax\protect\ifmmode}
\def\text{\RIfM@\expandafter\text@\else\expandafter\mbox\fi}
\let\nfss@text\text
\def\text@#1{\mathchoice
   {\textdef@\displaystyle\f@size{#1}}%
   {\textdef@\textstyle\tf@size{\firstchoice@false #1}}%
   {\textdef@\textstyle\sf@size{\firstchoice@false #1}}%
   {\textdef@\textstyle \ssf@size{\firstchoice@false #1}}%
   \glb@settings}

\def\textdef@#1#2#3{\hbox{{%
                    \everymath{#1}%
                    \let\f@size#2\selectfont
                    #3}}}
\newif\iffirstchoice@
\firstchoice@true
\def\Let@{\relax\iffalse{\fi\let\\=\cr\iffalse}\fi}%
\def\vspace@{\def\vspace##1{\crcr\noalign{\vskip##1\relax}}}%
\def\multilimits@{\bgroup\vspace@\Let@
 \baselineskip\fontdimen10 \scriptfont\tw@
 \advance\baselineskip\fontdimen12 \scriptfont\tw@
 \lineskip\thr@@\fontdimen8 \scriptfont\thr@@
 \lineskiplimit\lineskip
 \vbox\bgroup\ialign\bgroup\hfil$\m@th\scriptstyle{##}$\hfil\crcr}%
\def\Sb{_\multilimits@}%
\def\endSb{\crcr\egroup\egroup\egroup}%
\def\Sp{^\multilimits@}%

\newdimen\ex@
\def\rightarrowfill@#1{$#1\m@th\mathord-\mkern-6mu\cleaders
 \hbox{$#1\mkern-2mu\mathord-\mkern-2mu$}\hfill
 \mkern-6mu\mathord\rightarrow$}%
\def\leftarrowfill@#1{$#1\m@th\mathord\leftarrow\mkern-6mu\cleaders
 \hbox{$#1\mkern-2mu\mathord-\mkern-2mu$}\hfill\mkern-6mu\mathord-$}%
\def\leftrightarrowfill@#1{$#1\m@th\mathord\leftarrow
\mkern-6mu\cleaders
 \hbox{$#1\mkern-2mu\mathord-\mkern-2mu$}\hfill
 \mkern-6mu\mathord\rightarrow$}%
\def\overrightarrow{\mathpalette\overrightarrow@}%
\def\overrightarrow@#1#2{\vbox{\ialign{##\crcr\rightarrowfill@#1\crcr
 \noalign{\kern-\ex@\nointerlineskip}$\m@th\hfil#1#2\hfil$\crcr}}}%

\def\overleftarrow{\mathpalette\overleftarrow@}%
\def\overleftarrow@#1#2{\vbox{\ialign{##\crcr\leftarrowfill@#1\crcr
 \noalign{\kern-\ex@\nointerlineskip}$\m@th\hfil#1#2\hfil$\crcr}}}%
\def\overleftrightarrow{\mathpalette\overleftrightarrow@}%
\def\overleftrightarrow@#1#2{\vbox{\ialign{##\crcr
   \leftrightarrowfill@#1\crcr
 \noalign{\kern-\ex@\nointerlineskip}$\m@th\hfil#1#2\hfil$\crcr}}}%
\def\underrightarrow{\mathpalette\underrightarrow@}%
\def\underrightarrow@#1#2{\vtop{\ialign{##\crcr$\m@th\hfil#1#2\hfil
  $\crcr\noalign{\nointerlineskip}\rightarrowfill@#1\crcr}}}%

\def\underleftarrow{\mathpalette\underleftarrow@}%
\def\underleftarrow@#1#2{\vtop{\ialign{##\crcr$\m@th\hfil#1#2\hfil
  $\crcr\noalign{\nointerlineskip}\leftarrowfill@#1\crcr}}}%
\def\underleftrightarrow{\mathpalette\underleftrightarrow@}%
\def\underleftrightarrow@#1#2{\vtop{\ialign{##\crcr$\m@th
  \hfil#1#2\hfil$\crcr
 \noalign{\nointerlineskip}\leftrightarrowfill@#1\crcr}}}%

\def\qopnamewl@#1{\mathop{\operator@font#1}\nlimits@}
\let\nlimits@\displaylimits
\def\setboxz@h{\setbox\z@\hbox}

\def\varlim@#1#2{\mathop{\vtop{\ialign{##\crcr
 \hfil$#1\m@th\operator@font lim$\hfil\crcr
 \noalign{\nointerlineskip}#2#1\crcr
 \noalign{\nointerlineskip\kern-\ex@}\crcr}}}}

 \def\rightarrowfill@#1{\m@th\setboxz@h{$#1-$}\ht\z@\z@
  $#1\copy\z@\mkern-6mu\cleaders
  \hbox{$#1\mkern-2mu\box\z@\mkern-2mu$}\hfill
  \mkern-6mu\mathord\rightarrow$}
\def\leftarrowfill@#1{\m@th\setboxz@h{$#1-$}\ht\z@\z@
  $#1\mathord\leftarrow\mkern-6mu\cleaders
  \hbox{$#1\mkern-2mu\copy\z@\mkern-2mu$}\hfill
  \mkern-6mu\box\z@$}

\def\projlim{\qopnamewl@{proj\,lim}}
\def\injlim{\qopnamewl@{inj\,lim}}
\def\varinjlim{\mathpalette\varlim@\rightarrowfill@}
\def\varprojlim{\mathpalette\varlim@\leftarrowfill@}
\def\varliminf{\mathpalette\varliminf@{}}
\def\varliminf@#1{\mathop{\underline{\vrule\@depth.2\ex@\@width\z@
   \hbox{$#1\m@th\operator@font lim$}}}}
\def\varlimsup{\mathpalette\varlimsup@{}}
\def\varlimsup@#1{\mathop{\overline
  {\hbox{$#1\m@th\operator@font lim$}}}}

\def\stackunder#1#2{\mathrel{\mathop{#2}\limits_{#1}}}%
\begingroup \catcode `|=0 \catcode `[= 1
\catcode`]=2 \catcode `\{=12 \catcode `\}=12
\catcode`\\=12 
|gdef|@alignverbatim#1\end{align}[#1|end[align]]
|gdef|@salignverbatim#1\end{align*}[#1|end[align*]]

|gdef|@alignatverbatim#1\end{alignat}[#1|end[alignat]]
|gdef|@salignatverbatim#1\end{alignat*}[#1|end[alignat*]]

|gdef|@xalignatverbatim#1\end{xalignat}[#1|end[xalignat]]
|gdef|@sxalignatverbatim#1\end{xalignat*}[#1|end[xalignat*]]

|gdef|@gatherverbatim#1\end{gather}[#1|end[gather]]
|gdef|@sgatherverbatim#1\end{gather*}[#1|end[gather*]]

|gdef|@gatherverbatim#1\end{gather}[#1|end[gather]]
|gdef|@sgatherverbatim#1\end{gather*}[#1|end[gather*]]

|gdef|@multilineverbatim#1\end{multiline}[#1|end[multiline]]
|gdef|@smultilineverbatim#1\end{multiline*}[#1|end[multiline*]]

|gdef|@arraxverbatim#1\end{arrax}[#1|end[arrax]]
|gdef|@sarraxverbatim#1\end{arrax*}[#1|end[arrax*]]

|gdef|@tabulaxverbatim#1\end{tabulax}[#1|end[tabulax]]
|gdef|@stabulaxverbatim#1\end{tabulax*}[#1|end[tabulax*]]

|endgroup

\def\align{\@verbatim \frenchspacing\@vobeyspaces \@alignverbatim
You are using the "align" environment in a style in which it is not defined.}

\@namedef{align*}{\@verbatim\@salignverbatim
You are using the "align*" environment in a style in which it is not defined.}
\expandafter\let\csname endalign*\endcsname =\endtrivlist

\def\alignat{\@verbatim \frenchspacing\@vobeyspaces \@alignatverbatim
You are using the "alignat" environment in a style in which it is not defined.}

\@namedef{alignat*}{\@verbatim\@salignatverbatim
You are using the "alignat*" environment in a style in which it is not defined.}
\expandafter\let\csname endalignat*\endcsname =\endtrivlist

\def\xalignat{\@verbatim \frenchspacing\@vobeyspaces \@xalignatverbatim
You are using the "xalignat" environment in a style in which it is not defined.}

\@namedef{xalignat*}{\@verbatim\@sxalignatverbatim
You are using the "xalignat*" environment in a style in which it is not defined.}
\expandafter\let\csname endxalignat*\endcsname =\endtrivlist

\def\gather{\@verbatim \frenchspacing\@vobeyspaces \@gatherverbatim
You are using the "gather" environment in a style in which it is not defined.}

\@namedef{gather*}{\@verbatim\@sgatherverbatim
You are using the "gather*" environment in a style in which it is not defined.}
\expandafter\let\csname endgather*\endcsname =\endtrivlist

\def\multiline{\@verbatim \frenchspacing\@vobeyspaces \@multilineverbatim
You are using the "multiline" environment in a style in which it is not defined.}

\@namedef{multiline*}{\@verbatim\@smultilineverbatim
You are using the "multiline*" environment in a style in which it is not defined.}
\expandafter\let\csname endmultiline*\endcsname =\endtrivlist

\def\arrax{\@verbatim \frenchspacing\@vobeyspaces \@arraxverbatim
You are using a type of "array" construct that is only allowed in AmS-LaTeX.}

\def\tabulax{\@verbatim \frenchspacing\@vobeyspaces \@tabulaxverbatim
You are using a type of "tabular" construct that is only allowed in AmS-LaTeX.}

\@namedef{arrax*}{\@verbatim\@sarraxverbatim
You are using a type of "array*" construct that is only allowed in AmS-LaTeX.}
\expandafter\let\csname endarrax*\endcsname =\endtrivlist

\@namedef{tabulax*}{\@verbatim\@stabulaxverbatim
You are using a type of "tabular*" construct that is only allowed in AmS-LaTeX.}
\expandafter\let\csname endtabulax*\endcsname =\endtrivlist

\def\@@eqncr{\let\@tempa\relax
    \ifcase\@eqcnt \def\@tempa{& & &}\or \def\@tempa{& &}%
      \else \def\@tempa{&}\fi
     \@tempa
     \if@eqnsw
        \iftag@
           \@taggnum
        \else
           \@eqnnum\stepcounter{equation}%
        \fi
     \fi
     \global\tag@false
     \global\@eqnswtrue
     \global\@eqcnt\z@\cr}

 \def\endequation{%
     \ifmmode\ifinner 
      \iftag@
        \addtocounter{equation}{-1} 
        $\hfil
           \displaywidth\linewidth\@taggnum\egroup \endtrivlist
        \global\tag@false
        \global\@ignoretrue   
      \else
        $\hfil
           \displaywidth\linewidth\@eqnnum\egroup \endtrivlist
        \global\tag@false
        \global\@ignoretrue 
      \fi
     \else   
      \iftag@
        \addtocounter{equation}{-1} 
        \eqno \hbox{\@taggnum}
        \global\tag@false%
        $$\global\@ignoretrue
      \else
        \eqno \hbox{\@eqnnum}
        $$\global\@ignoretrue
      \fi
     \fi\fi
 } 

 \newif\iftag@ \tag@false
 
 \def\tag{\@ifnextchar*{\@tagstar}{\@tag}}
 \def\@tag#1{%
     \global\tag@true
     \global\def\@taggnum{(#1)}}
 \def\@tagstar*#1{%
     \global\tag@true
     \global\def\@taggnum{#1}%
}

\makeatother

\begin{document}

\begin{center}
\textbf{DESCRIPTION OF THE TWO-NUCLEON SYSTEM ON THE BASIS OF THE BARGMANN
REPRESENTATION OF THE }$\mathbf{S}$\textbf{\ MATRIX}

\textbf{\ }\\[0pt]
\textbf{V. A. Babenko and N. M. Petrov \footnote{{\normalsize
\mbox{E-mail address:
pet@online.com.ua}}}}\\[0pt]
\textit{Bogolyubov Institute for Theoretical Physics,}

\textit{National Academy of Sciences of Ukraine, Kiev, Ukraine}
\end{center}

\vspace{1pt}

\vspace{1pt}

\noindent For the effective-range function $k\cot \delta $, a pole
approximation that involves a small number of parameters is derived on the
basis of the Bargmann representation of the $S$ matrix. The parameters of
this representation, which have a clear physical meaning, are related to the
parameters of the Bargmann $S$ matrix by simple equations. By using a
polynomial least-squares fit to the function $k\cot \delta $ at low
energies, the triplet low-energy parameters of neutron-proton scattering are
obtained for the latest experimental data of Arndt et al. on phase shifts.
The results are $a_{t}=5.4030\,$fm, $r_{t}=1.7494\,$fm, and $v_{2}=0.163\,$fm%
$^{3}$. With allowance for the values found for the low-energy scattering
parameters and for the pole parameter, the pole approximation of the
function $k\cot \delta $ provides an excellent description of the triplet
phase shift for neutron-proton scattering over a wide energy range ($T_{%
\text{lab}}\lesssim 1000\,$MeV), substantially improving the description at
low energies as well. For the experimental phase shifts of Arndt et al., the
triplet shape parameters $v_{n}$ of the effective-range expansion are
obtained by using the pole approximation. The description of the phase shift
by means of the effective-range expansion featuring values found for the
low-energy scattering parameters proves to be fairly accurate over a broad
energy region extending to energy values approximately equal to the energy
at which this phase shift changes sign, this being indicative of a high
accuracy and a considerable value of the effective-range expansion in
describing experimental data on nucleon-nucleon scattering. The properties
of the deuteron that were calculated by using various approximations of the
effective-range function comply well with their experimental values.

\vspace{1pt}

\vspace{1pt}\newpage

\begin{center}
\vspace{1pt}1. INTRODUCTION
\end{center}

\vspace{1pt}\vspace{1pt}

The $S$ matrix \lbrack 1, 2\rbrack\ is a fundamental quantity in scattering
theory. At a fixed value of the orbital angular momentum $\ell $, the $S$
matrix is a function of the wave number $k$ related to the energy $E$ of two
colliding particles in the c.m. frame by the equation $E=\hbar
^{2}k^{2}/2m\, $, where $m$ is the reduced mass of the system and $\hbar $\
is the Plank constant. In the following, we restrict our consideration to
the case of zero orbital angular momentum (we omit the index $\ell =0$ for
the sake of simplicity). The fact that a unified description of both
scattering and bound states in two-particle systems can be constructed on
the basis of the respective $S$ matrix is an important consequence of the
analytic properties of this matrix. In general, the analytic properties of
the $S$ matrix in the $k$ plane are rather complicated; however, a number of
important results can be obtained in considering specific physical systems
by using an $S$ matrix whose analytic properties are simple. In the $k$
plane, the $S$ matrix satisfies the relation \lbrack 1\rbrack
\begin{equation}
S^{\ast }\left( k^{\ast }\right) \cdot S\left( k\right) =1\,.  \tag{1}
\end{equation}
At real $k$ values, this relation coincides with the unitarity condition.

\vspace{1pt}

In potential-scattering theory, the $S$ matrix $S\left( k\right) $ is
expressed in terms of the Jost function $F\left( k\right) $ as \lbrack 3,
4\rbrack
\begin{equation}
S\left( k\right) =\frac{F\left( -k\right) }{F\left( k\right) }\,.  \tag{2}
\end{equation}
At complex values of $k$, the Jost function $F\left( k\right) $ satisfies
the condition
\begin{equation}
F^{\ast }\left( -k^{\ast }\right) =F\left( k\right) \,.  \tag{3}
\end{equation}
The $S$-matrix property (1) immediately follows from this condition with
allowance for (2). The behavior of the Jost function at high energies is
determined by the relation
\begin{equation}
\stackunder{k\rightarrow \infty }{\lim }F\left( k\right) =1\,.  \tag{4}
\end{equation}
The behavior of the $S$ matrix at high energies directly follows from (2)
and (4):
\begin{equation}
\stackunder{k\rightarrow \infty }{\lim }S\left( k\right) =1\,.  \tag{5}
\end{equation}

\vspace{1pt}

In the case of elastic scattering, the $S$ matrix can be expressed in terms
of the phase shift $\delta \left( k\right) $ as
\begin{equation}
S\left( k\right) =e^{2i\delta \left( k\right) }\,.  \tag{6}
\end{equation}
Using relation (6) and considering that the $S$ matrix possesses the
symmetry property \lbrack 1\rbrack
\begin{equation}
S\left( -k\right) =S^{-1}\left( k\right) \,,  \tag{7}
\end{equation}
one can easily see that, at real values of $k$, the phase shift $\delta
\left( k\right) $ is an odd function of $k$; that is,
\begin{equation}
\delta \left( -k\right) =-\delta \left( k\right) \,.  \tag{8}
\end{equation}
From relations (5) and (6), one readily establishes the asymptotic behavior
of the phase shift at high energies:
\begin{equation}
\stackunder{k\rightarrow \infty }{\lim }\delta \left( k\right) =0\,.  \tag{9}
\end{equation}

\vspace{1pt}

In practice, it is very useful to introduce the effective-range function
\begin{equation}
k\cot \delta =ik\frac{S\left( k\right) +1}{S\left( k\right) -1}\,.  \tag{10}
\end{equation}
In the physical region, the function $k\cot \delta $ is real by virtue of
the unitarity of the $S$ matrix, while, in the $k$ plane, this function is
analytic everywhere, with the exception of the points where $S\left(
k\right) =1$, i.e. the points where it has poles. The introduction of the
function $k\cot \delta $ simplifies the investigation of the analytic
properties of the $S$ matrix. It follows from (8) that the function $k\cot
\delta $ is an even function of $k$.

\vspace{1pt}

In studying nucleon-nucleon ($NN$) scattering at low energies, it is useful
to employ the well-known effective-range expansion
\begin{equation}
k\cot \delta =-\frac{1}{a}+\frac{1}{2}%
r_{0}k^{2}+v_{2}k^{4}+v_{3}k^{6}+v_{4}k^{8}+\ldots \,,  \tag{11}
\end{equation}
which involves only even powers of $k$. The effective-range approximation
corresponds to retaining only the first two terms in (11). In expansion
(11), the quantities $a$ and $r_{0}$ are, respectively, the scattering
length and the effective range, while the quantities $v_{n}$ are related to
the potential shape. It should be noted that the quantity $r_{0}$ is a
measure of the effective interaction region. In the case of nucleon-nucleon
interaction, the effective range $r_{0}$ is approximately equal to the range
of nuclear forces ($R$). If, however, one deals with the doublet
neutron-deuteron ($nd$) interaction, the effective range is anomalously
large ($r_{0}\sim 500\,$fm), so that the statement that it is approximately
equal to the range of nuclear forces ($R\sim 2\,$fm) does not hold here.
This is because, for $nd$ interaction, the function $k\cot \delta $ has a
pole in the vicinity of the point $k^{2}=0$ owing to the existence of a
low-energy virtual triton state \lbrack 5\rbrack . \ \ \ \

\vspace{1pt}

\vspace{1pt}In a number of studies (see, for example, \lbrack 6--8\rbrack ),
the authors approximated the $S$ matrix by rational functions, this leading
to so-called Bargmann potentials \lbrack 4, 9\rbrack . The use of a Bargmann
$S$ matrix makes it possible to find explicit solutions to the direct and
inverse scattering problems. Of particular importance are special cases
where the Bargmann $S$ matrix is determined by a small number of parameters
such that physical observables are directly expressed in terms of these
parameters. In \lbrack 5\rbrack , we showed that the use of the Bargmann $S$
matrix corresponding to the presence of two states in the system being
considered leads to the pole structure of the effective-range function (van
Oers-Seagrave formula \lbrack 10\rbrack ). At low energies, this structure
makes it possible to describe well doublet $nd$ scattering on the basis of
parameters that characterize the bound and the virtual state of the triton.

\vspace{1pt}

In the present study, we use the Bargmann representation of the $S$ matrix
to construct the pole approximation of the effective-range function, bearing
in mind that this pole approximation is optimal for describing
nucleon-nucleon scattering and that it involves a small number of
parameters. The pole-approximation parameters have a clear physical meaning,
and the values found for these parameters from an analysis of low-energy
experimental data make it possible to obtain an excellent unified
description of a bound state in the two-nucleon system (deuteron) and the
triplet phase shift for neutron-proton scattering over a very wide energy
range ($0-1000\,$MeV).

\vspace{1pt}

\vspace{1pt}

\begin{center}
2. BARGMANN REPRESENTATION OF THE $S$ MATRIX AND POLE APPROXIMATION OF THE
EFFECTIVE-RANGE FUNCTION\vspace{1pt}
\end{center}

\vspace{1pt}\vspace{1pt}

In order to investigate the interaction in the system of two particles, we
will use the corresponding Bargmann $S$ matrix \lbrack 4, 9\rbrack , which
possesses simple analytic properties in the $k$ plane. In this case, the
Jost function $F\left( k\right) $ can be chosen in the simplest way in the
form of a rational function, 
\begin{equation}
F\left( k\right) =\stackrel{N}{\stackunder{n=1}{\Pi }}\frac{k-i\alpha _{n}}{%
k+i\lambda _{n}}\,.  \tag{12}
\end{equation}
It has $N$ simple zeros at the points $k=i\alpha _{n}$ and $N$ simple poles
at the points $k=-i\lambda _{n}$ and exhibits the correct asymptotic
behavior (4) at high energies.

\vspace{1pt}

As a simple example of a Bargmann $S$ matrix, one can consider the truncated 
$S$ matrix corresponding to the well-known explicitly solvable Hulth\'{e}n
potential 
\begin{equation}
V\left( r\right) =-V_{0}\left( e^{r/R}-1\right) ^{-1}\,,  \tag{13}
\end{equation}
for which we have 
\begin{equation}
\alpha _{n}=\frac{g-n^{2}}{2nR}\,,\,\,\,\,\,\lambda _{n}=\frac{n}{2R}\,, 
\tag{14}
\end{equation}
where 
\begin{equation}
g=\frac{2m}{\hbar ^{2}}V_{0}R^{2}\,.  \tag{15}
\end{equation}

\vspace{1pt}

The Bargmann scattering matrix 
\begin{equation}
S\left( k\right) =\stackrel{N}{\stackunder{n=1}{\Pi }}\frac{k+i\alpha _{n}}{%
k-i\alpha _{n}}\frac{k+i\lambda _{n}}{k-i\lambda _{n}}\,,  \tag{16}
\end{equation}
which corresponds to the Jost function (12), has $N$ simple ``physical''
poles at $k=i\alpha _{n}$ in the $k$ plane. The poles lying on the imaginary
axis in the upper half-plane ($\alpha _{n}>0$) correspond to bound states,
while the poles in the lower half-plane correspond to resonances or virtual
states. The resonances in the lower half-plane are grouped in pairs
symmetric with respect to the imaginary axis, while the virtual states are
on the imaginary axis ($\alpha _{n}<0$). In addition to the physical poles
corresponding to resonances and bound and virtual states, the Bargmann $S$
matrix (16) has $N$ so-called ``redundant'' poles at $k=i\lambda _{n}$ in
the upper half-plane, which do not correspond to any physical states. The
presence, along with physical poles, of the same number of redundant poles
serves as some kind of a compensation factor, ensuring the correct
asymptotic behavior \lbrack see Eq. (4)\rbrack\ of the Jost function at
infinity. \ \ 

\vspace{1pt}

With allowance for Eqs. (2) and (3), the effective-range function (10) at
real values of $k$ can be expressed in terms of the real and the imaginary
part of the Jost function as 
\begin{equation}
k\cot \delta =-k\frac{\func{Re}F\left( k\right) }{\func{Im}F\left( k\right) }%
\,.  \tag{17}
\end{equation}
\ From the Bargmann representation (16) of the $S$ matrix, it immediately
follows that the effective-range function $k\cot \delta $ can then be
written in the form of a rational function of $k^{2}$, 
\begin{equation}
k\cot \delta =\frac{P_{N}\left( k^{2}\right) }{Q_{N-1}\left( k^{2}\right) }%
\,,  \tag{18}
\end{equation}
where the degrees of the polynomials $P_{N}\left( k^{2}\right) $ and $%
Q_{N-1}\left( k^{2}\right) $ of $k^{2}$ are $N$ and $N-1$, respectively, and
where the coefficients in these polynomials are completely determined by the
quantities $\alpha _{n}$ and $\lambda _{n}$.

\vspace{1pt}

\vspace{1pt}The representation in (18) for the function $k\cot \delta $ can
be considered as a Pad\'{e} approximation of this function \lbrack
11\rbrack\ if the expansion of the function $P_{N}\left( k^{2}\right)
/Q_{N-1}\left( k^{2}\right) \equiv \lbrack N/\left( N-1\right) \rbrack $\ in
a Taylor-Maclaurin series coincides with the effective-range expansion (11)
up to terms of order $2N-1$. In other words, the coefficients of $%
1,\,k^{2},\,\ldots \,,\,k^{2\left( 2N-1\right) }$ in the Taylor expansion of
the function $\lbrack N/\left( N-1\right) \rbrack $ must coincide with the
corresponding coefficients in the series in (11). The Pad\'{e} approximation
method was used by various authors to study nucleon-nucleon scattering (see,
for example, \lbrack 12--15\rbrack ). It is of importance in our case that,
from the Bargmann representation (16) of the $S$ matrix, it automatically
follows that the degree of the polynomial $P$ in the numerator of the
Pad\'{e} approximant (18) must be greater by unity than the degree of the
polynomial $Q$ in its denominator, this being in accord with the condition
of the theorem \lbrack 12\rbrack\ on the solvability of the inverse
scattering problem. In addition, one can see that, in the majority of cases,
the condition $L=M+1$ for the Pad\'{e} approximant $\lbrack L/M\rbrack $ of
the function $k\cot \delta $ is optimal for specific fits \lbrack
13--15\rbrack .

\vspace{1pt}

Let us now consider in detail the important particular case where the
Bargmann $S$ matrix corresponds to the presence of two physical states ($N=2$%
) in the system. This is so, for example, in the doublet scattering of a
neutron on a deuteron \lbrack 5\rbrack , in which case there are two triton
states in the system, a bound and a virtual one. The scattering matrix
corresponding to the presence of two states in the system can also be used
to describe triplet neutron-proton scattering. Since the second state has
not yet been observed experimentally in that case, the energy of this state
must be considerably higher than the deuteron binding energy. If, in this
case, the second state is a bound state of two nucleons, such a situation
corresponds to the phenomenology of nodes that is described by a short-range
deep attractive potential involving forbidden states \lbrack 16--18\rbrack .

\vspace{1pt}

If the system has two physical states, the rational Jost function and the
Bargmann $S$ matrix corresponding to it are given by 
\begin{equation}
F\left( k\right) =\frac{k-i\alpha }{k+i\lambda }\frac{k-i\beta }{k+i\mu }\,,
\tag{19}
\end{equation}
\begin{equation}
S\left( k\right) =\frac{k+i\alpha }{k-i\alpha }\frac{k+i\beta }{k-i\beta }%
\frac{k+i\lambda }{k-i\lambda }\frac{k+i\mu }{k-i\mu }\,.  \tag{20}
\end{equation}
The first and the second factors in the $S$-matrix representation (20)
correspond to either bound or virtual states of the system, while the third
and the fourth factors correspond to redundant poles of the $S$ matrix. The
negative energies of the bound and virtual states of the system are 
\begin{equation}
E_{\alpha }=-\frac{\hbar ^{2}\alpha ^{2}}{2m}\,,  \tag{21}
\end{equation}
\begin{equation}
E_{\beta }=-\frac{\hbar ^{2}\beta ^{2}}{2m}\,.  \tag{22}
\end{equation}

\vspace{1pt}

\vspace{1pt}The expression describing the effective-range function $k\cot
\delta $ and corresponding to the presence of two states in the system
immediately follows from (17) and (19). The result is 
\begin{equation}
k\cot \delta =-\frac{1}{a}\,\frac{1-c_{2}k^{2}+c_{4}k^{4}}{1+Dk^{2}}\,, 
\tag{23}
\end{equation}
where, for the sake of convenience, we have explicitly isolated the
scattering length $a$. The parameters appearing in the pole representation
(23) of the function $k\cot \delta $ and the parameters $\alpha $, $\beta $, 
$\lambda $, and $\mu $ of the Bargmann $S$ matrix (20) are related by the
equations 
\begin{equation}
a=\frac{1}{\alpha }+\frac{1}{\beta }+\frac{1}{\lambda }+\frac{1}{\mu }\,, 
\tag{24}
\end{equation}
\begin{equation}
c_{2}=\frac{1}{\alpha \beta }+\frac{1}{\alpha \lambda }+\frac{1}{\alpha \mu }%
+\frac{1}{\beta \lambda }+\frac{1}{\beta \mu }+\frac{1}{\lambda \mu }\,, 
\tag{25}
\end{equation}
\begin{equation}
c_{4}=\frac{1}{\alpha \beta \lambda \mu }\,,  \tag{26}
\end{equation}
\begin{equation}
D=-\frac{1}{a}\,\frac{\alpha +\beta +\lambda +\mu }{\alpha \beta \lambda \mu 
}\,.  \tag{27}
\end{equation}
The parameter $D$ appearing in (23) determines the pole of the
effective-range function $k_{0}^{2}$: 
\begin{equation}
k_{0}^{2}=-\frac{1}{D}\,.  \tag{28}
\end{equation}
\ \vspace{1pt}

\vspace{1pt}

\vspace{1pt}Expression (23), which was obtained for the function $k\cot
\delta $ by the above method from the Bargmann representation of the $S$
matrix, is a one-pole representation and involves four independent
parameters. It can easily be shown that, apart from the form of
presentation, the pole expression (23) coincides with the well-known
empirical formulas that were given by van Oears and Seagrave \lbrack
10\rbrack\ and by Cini, Fubini, and Stanghellini \lbrack 19\rbrack\ and
which are applied to describe, respectively, neutron-deuteron and
nucleon-nucleon scattering. It is worth noting that various forms of the
pole structure of the effective-range function $k\cot \delta $ have been
repeatedly discussed for a rather long time and have been successfully used
to describe neutron-deuteron \lbrack 5, 10, 20--30\rbrack\ and
nucleon-nucleon interactions \lbrack 19, 31--35\rbrack . In the majority of
cases, however, the formulas for $k\cot \delta $ were obtained empirically.
But in our case, the pole formula for the function $k\cot \delta $
immediately follows from the Bargmann $S$ matrix, which possesses simple
properties and which takes into account basic physical properties of the
interaction in the system being considered.

\vspace{1pt}

In \lbrack 5\rbrack , we showed that, in the case of neutron-deuteron
interaction, the presence of the pole in the function $k\cot \delta $ is a
direct consequence of the existence of a virtual triton state at a low
energy. In the present study, the effective-range function having a pole
structure will be used to describe neutron-proton interaction in the triplet
(t) spin state $^{3}S_{1}$. In this case, the neutron-proton system has one
bound state (deuteron), the scattering of a neutron on a proton at low
energies (up to energies of about $10\,$MeV) being well described in the
effective-range approximation. In this connection, it is convenient to
recast the pole-approximation formula (23) into the form 
\begin{equation}
k\cot \delta =-\frac{1}{a_{t}}+\frac{1}{2}r_{t}k^{2}+\frac{v_{2}k^{4}}{%
1+Dk^{2}}\,,  \tag{29}
\end{equation}
where $a_{t}$ and $r_{t}$ are, respectively, the scattering length and the
effective range, while the parameter $v_{2}$ determines the dimensionless
shape parameter $P_{t}$, which is widely used in the literature, according
to the relation 
\begin{equation}
P_{t}=v_{2}/r_{t}^{3}\,.  \tag{30}
\end{equation}
We also note that, in the literature, definition of the dimensionless shape
parameter $P_{t}$ frequently differs from the one in Eq. (30) by sign. But
our definition here is more convenient for future discussion.

\vspace{1pt}

\vspace{1pt}The first two terms of the representation in (29) correspond to
the effective-range approximation, while the last, pole, term describes the
deviation from this approximation. The presence of this pole term makes it
possible to improve, with the aid of only two additional parameters, the
description of the phase shift significantly and to extend the range of
applicability of this description greatly. Thus, the pole approximation
(29), which was derived from the Bargmann representation of the $S$ matrix,
is a direct generalization of the effective-range approximation to the case
where there are two physical states in the system. We note that the form
(29) of the pole formula is more convenient for describing nucleon-nucleon
scattering than that in (23), since, in (29), the low-energy scattering
parameters $a_{t}$, $r_{t}$, and $v_{2}$ are isolated explicitly.

\vspace{1pt}

\vspace{1pt}The parameters $a_{t}$ and $D$ appearing in (29) are related to
the parameters of the Bargmann $S$ matrix (20) by Eqs. (24) and (27), while
the effective range $r_{t}$ and the shape parameter $v_{2}$ are given by 
\begin{equation}
r_{t}=\frac{2}{a_{t}}\left( D+c_{2}\right) \,,  \tag{31}
\end{equation}
\begin{equation}
v_{2}=-\left( \frac{1}{2}Dr_{t}+\frac{c_{4}}{a_{t}}\right) \,,  \tag{32}
\end{equation}
where $c_{2}$ and $c_{4}$ are related to the $S$-matrix parameters by Eqs.
(25) and (26).

\vspace{1pt}\vspace{1pt}

Let us now consider the specific case where the $S$ matrix in (20) for the
case of two states reduces to the $S$ matrix for one state, 
\begin{equation}
S\left( k\right) =\frac{k+i\alpha }{k-i\alpha }\frac{k+i\lambda }{k-i\lambda 
}\,.  \tag{33}
\end{equation}
In this case, the second state goes to infinity: $\beta \rightarrow \infty $
and $\mu \rightarrow \infty $. It can easily be seen that the coefficients $%
D $ and $v_{2}$ then vanish in formula (29), which reduces, after that, to
the effective-range-approximation formula 
\begin{equation}
k\cot \delta =-\frac{1}{a_{t}}+\frac{1}{2}r_{t}k^{2}\,,  \tag{34}
\end{equation}
where the scattering length $a_{t}$ and the effective range $r_{t}$ are
given by 
\begin{equation}
a_{t}=\frac{1}{\alpha }+\frac{1}{\lambda }\,,  \tag{35}
\end{equation}
\begin{equation}
r_{t}=\frac{2}{\alpha +\lambda }\,.  \tag{36}
\end{equation}

\vspace{1pt}

The dimensionless asymptotic normalization factor $C_{d}$ characterizing the
bound state of the two-nucleon system (deuteron) can be expressed in terms
of the residue of the $S$ matrix at the pole $k=i\alpha $ as 
\begin{equation}
C_{d}^{2}=\frac{i}{2\alpha }\stackunder{k=i\alpha }{\limfunc{Res}}S\left(
k\right) \,.  \tag{37}
\end{equation}
In the case of the $S$ matrix for one state, $C_{d}$ is given by 
\begin{equation}
C_{d}^{2}=\frac{\lambda +\alpha }{\lambda -\alpha }\,.  \tag{38}
\end{equation}
Upon expressing $\lambda $ in terms of $\alpha $ and $C_{d}^{2}$, formulas
(35) and (36) for the scattering length $a_{t}$ and the effective range $%
r_{t}$, respectively, can be recast into the form \lbrack 36\rbrack 
\begin{equation}
a_{t}=\frac{2}{\alpha }\frac{C_{d}^{2}}{1+C_{d}^{2}}\,,  \tag{39}
\end{equation}
\begin{equation}
r_{t}=\frac{1}{\alpha }\left( 1-\frac{1}{C_{d}^{2}}\right) \,.  \tag{40}
\end{equation}
Thus, formulas (39) and (40) give explicit expressions for the low-energy
scattering parameters $a_{t}$ and $r_{t}$ in terms of parameters that
characterize the bound state of the two-nucleon system (deuteron) in the
case where the interaction in the system is described by the $S$ matrix for
one state. The inverse statement is also valid --- namely, the parameters
characterizing the deuteron can be expressed, in this case, in terms of the
scattering parameters $a_{t}$ and $r_{t}$ as 
\begin{equation}
\alpha =\frac{1}{r_{t}}\left[ 1-\left( 1-\frac{2r_{t}}{a_{t}}\right)
^{1/2}\right] \,,  \tag{41}
\end{equation}
\begin{equation}
C_{d}^{2}=\frac{1}{\left( 1-2r_{t}/a_{t}\right) ^{1/2}}\,.  \tag{42}
\end{equation}

\vspace{1pt}

The case where, at finite values of $\beta $ and $\mu $, the coefficient $D$
vanishes, while the parameter $v_{2}$ differs from zero is yet another
important specific case of formula (29). This is so if the system has a
virtual state whose wave number $\beta $ is given by 
\begin{equation}
\beta =-\left( \alpha +\lambda +\mu \right) \,.  \tag{43}
\end{equation}
It can be seen from (27) that, in this case, the parameter $D$ is equal to
zero, while expression (29) for the effective-range function reduces to the
expression corresponding to the shape-parameter approximation, 
\begin{equation}
k\cot \delta =-\frac{1}{a_{t}}+\frac{1}{2}r_{t}k^{2}+v_{2}k^{4}\,,  \tag{44}
\end{equation}
where 
\begin{equation}
r_{t}=2\frac{c_{2}}{a_{t}}\,,  \tag{45}
\end{equation}
\begin{equation}
v_{2}=P_{t}r_{t}^{3}=-\frac{c_{4}}{a_{t}}\,,  \tag{46}
\end{equation}
with $a_{t}$, $c_{2}$, and $c_{4}$ being given by (24)--(26). The
effective-range approximation (34) and the shape-parameter approximation
(44) are among the most popular and important methods for parametrizing data
on nucleon-nucleon scattering at low energies \lbrack 37, 38\rbrack .

\vspace{1pt}

\vspace{1pt}In the triplet spin state, the neutron-proton scattering length $%
a_{t}$ is positive. It follows that, in the approximation specified by Eq.
(44), the sign of the shape parameter $P_{t}$ is determined by the sign of
the parameter $c_{4}$ in accordance with (46). Since the redundant poles $%
k=i\lambda $ and $k=i\mu $ of the $S$ matrix lie in the upper half-plane and
since $\alpha >0$ and $\beta <0$, it follows from (26) that $c_{4}$ is
negative. Therefore, the dimensionless shape parameter $P_{t}$ is positive
in this approximation: 
\begin{equation}
P_{t}>0\,.  \tag{47}
\end{equation}

\vspace{1pt}

\vspace{1pt}Thus, we have shown that the shape-parameter approximation (44)
directly follows from the Bargmann representation of the $S$ matrix in the
case where the system has two physical states of which at least one is
virtual. If the system has two virtual states ($\alpha <0$, $\beta <0$), the
shape-parameter approximation (44) follows from the Bargmann representation
of the $S$ matrix under the condition 
\begin{equation}
\left( \alpha +\beta \right) =-\left( \lambda +\mu \right) \,  \tag{48}
\end{equation}
and corresponds to the description of the neutron-proton interaction in the
singlet spin state.

\vspace{1pt}\vspace{1pt}

\begin{center}
\vspace{1pt}3. CHOICE OF PARAMETERS AND DESCRIPTION OF SCATTERING IN THE
TWO-NUCLEON SYSTEM
\end{center}

\vspace{1pt}\vspace{1pt}

\vspace{1pt}An investigation of the phase shift $\delta $ as function of
energy plays a fundamental role in an analysis of data on nucleon-nucleon
scattering. A traditional way to study this dependence at low energies
consists in applying the effective-range approximation (34) and the
shape-parameter approximation (44) \lbrack 37, 38\rbrack . In order to
determine the parameters $a_{t}$, $r_{t}$, and $P_{t}$, one employs
experimental data on scattering and also the experimental value of the
deuteron binding energy $\varepsilon _{d}$. It was shown in \lbrack
38\rbrack\ that, in this case, the determination of the scattering
parameters $a_{t}$, $r_{t}$, and $P_{t}$ involved ambiguities, and this is
at odds with the meaning of the effective-range expansion (11). The
ambiguity in the determination of the low-energy scattering parameters was
due primarily to an insufficient accuracy of experimental data at low
energies. At the present time, the accuracy of experimental data is such
that the ambiguity in the determination of the scattering length $a_{t}$,
the effective range $r_{t}$, and the shape parameter $P_{t}$ is quite
removable.

\vspace{1pt}

By using the least squares method to construct a polynomial fit to the
triplet function $k\cot \delta $ at low energies ($T_{\text{lab}}\lesssim
10\,\,$MeV) and relying on the latest experimental data of the VPI/GWU group
of Arndt et al. on neutron-proton phase shifts \lbrack 39\rbrack , we obtain
the following values of the triplet low-energy scattering parameters $a_{t}$%
, $r_{t}$, and $\,v_{2}$: 
\begin{equation}
a_{t}=5.4030\,\text{fm\thinspace },  \tag{49}
\end{equation}
\begin{equation}
r_{t}=1.7494\,\text{fm\thinspace },  \tag{50}
\end{equation}
\begin{equation}
v_{2}=0.163\,\text{fm}^{3}\,.  \tag{51}
\end{equation}
The resulting dimensionless shape parameter, $P_{t}=v_{2}/r_{t}^{3}=0.0304$,
proved to be positive. This is in agreement with the estimate in (47), which
was obtained on the basis of the Bargmann representation of the $S$ matrix.

\vspace{1pt}

In order to describe triplet scattering and the bound state in the
two-nucleon system, we will use the pole approximation of the
effective-range function (29). In doing this, we set the low-energy
parameters $a_{t}$, $r_{t}$, and $v_{2}$ to the values in (49)--(51) and
choose the value of the pole parameter $D$ to be 
\begin{equation}
D=-0.225526\,\text{fm}^{2}\,.  \tag{52}
\end{equation}
With allowance for (28), this corresponds to the laboratory energy 
\begin{equation}
T_{0}=368.026\,\text{MeV}\,,  \tag{53}
\end{equation}
at which the experimental value of the triplet phase shift \lbrack
39\rbrack\ changes sign. Thus, it can be seen, that, if the pole parameter $%
D $ is fixed by using the experimental point $\delta \left( T_{0}\right) =0$%
, the quantity $D$ can be considered as an experimental parameter directly
determined to a rather high degree of accuracy.

\vspace{1pt}

\vspace{1pt}If one uses the parameter values in (49)--(52), then, as can be
seen from the results of phase-shift calculations in Table 1 (the pole
approximation P1), the pole approximation (29) describes the experimental
triplet phase shift \lbrack 39\rbrack\ up to laboratory energies of about $%
500\,$MeV with an absolute error not exceeding $0.5^{\circ }$. For the sake
of comparison, the results obtained by calculating the phase shift within
the effective-range approximation (ER) and within the shape-parameter
approximation (SP) with the low-energy parameters set to the values in
(49)--(51) are also presented in Table 1. As might have been expected, the
effective-range approximation (34) describes well the phase shift only in
the region of very low energies ($T_{\text{lab}}\lesssim 10\,$MeV). The
shape-parameter approximation (44) describes well the experimental phase
shift up to the energies of about $50\,$MeV (to within about $0.5^{\circ }$%
). At the same time, the pole approximation (29) provides an excellent
description of the phase shift within a broad energy range, considerably
improving the description in the low-energy region --- for example, the
experimental phase shift in the energy range $T_{\text{lab}}\leqslant 40\,$%
MeV is described by formula (29) with an error not exceeding $0.01^{\circ }$%
, i.e. with better accuracy than accuracy of the experimental data (which is
typically $\gtrsim 0.01^{\circ }$). In the above energy range, one can
therefore consider the parametrization of the phase shift by the pole
formula (29) as an alternative to the data of a partial-wave analysis (PWA).
As a matter of fact, the approximation specified by Eq. (29) with the
parameters given in (49)--(52) provides an excellent description of the
phase shift in much wider energy interval than that which was used to find
the parameters of this approximation. As can be seen from Fig. 1, the
accuracy of the description of the phase shift undergoes virtually no
deterioration up to an energy of $900\,$MeV, being $0.6^{\circ }$ for
laboratory energies of $T_{\text{lab}}\leqslant 900\,$MeV. Along with the
phase shift corresponding to the pole approximation (29), the phase shift
calculated within the effective-range approximation (34) and that calculated
within the shape-parameter approximation (44) are also displayed in Fig. 1
for the sake of comparison.

\vspace{1pt}

By slightly varying the shape parameter $v_{2}$ and leaving the parameters $%
a_{t}$, $r_{t}$, and $D$ unchanged, one can improve further the quality of
description of the phase shift in the energy range $T_{lab}\lesssim 400\,$%
MeV. The minimum absolute error of description of the phase shift in this
energy range is attained at 
\begin{equation}
v_{2}=0.168\,\text{fm}^{3}  \tag{54}
\end{equation}
and is about $0.1^{\circ }$, as can be seen in Table 1 (pole approximation
P2). Thus, the pole approximation (29) makes it possible to describe, by
using a small number of parameters, the phase shift over a wide energy range
to a precision close to that in determining experimental data.

\vspace{1pt}\vspace{1pt}\vspace{1pt}\vspace{1pt}

It should be noted that the value of the pole parameter $D$ is very well
determined by the experimental phase shift even at low energies and is close
to the above value in (52). This can be shown most clearly by analyzing the
dependence of the absolute error of the description of the phase shift, 
\begin{equation}
\Delta =\stackunder{0\leqslant T\leqslant T_{\max }}{\max }\left| \delta _{%
\text{expt}}\left( T\right) -\delta _{\text{theor}}\left( T\right) \right|
\,,  \tag{55}
\end{equation}
on the parameter $D$ in a given range $0\leqslant T\leqslant T_{\max }$. The
dependence of the accuracy $\Delta $ on the parameter $D$ for the energy
interval $0\leqslant T\leqslant 20\,$MeV$\,$ is displayed in Fig. 2. The
phase shift in this case is described by the pole approximation (29) at
fixed values of the low-energy parameters $a_{t}$,$\,\,r_{t}$, and $v_{2}$\
(49)--(51), and at varying value of the pole parameter $D$.\ As can be seen
from Fig. 2, the minimal error is achieved at negative values of $D$ that
are close to the value in (52), the dependence $\Delta \left( D\right) $
having quite a sharp character in the vicinity of the minimum, so that any
significant deviation of the parameter $D$ from the point of the minimum
leads to a considerable deterioration of the quality of phase-shift
description. We note that the value of $D=0$ corresponds to the
shape-parameter approximation and that, for $D\rightarrow \infty $, we
obtain the effective-range approximation. The calculations show that, at the
value of $D=-0.2147\,$fm$^{2}$, which corresponds to the minimal deviation $%
\Delta $, the absolute error of the description of the phase shift on the
basis of (29) does not exceed $0.006^{\circ }$ in the energy range being
considered. At the same time, $\Delta $ takes values of $1.16^{\circ }$ and $%
0.064^{\circ }$ for, respectively, the effective-range and the
shape-parameter approximation. Thus, we see that, even at low energies, the
quality of the description of the phase shift by the pole formula is an
order of magnitude higher than the quality of the description within the
shape-parameter approximation, the value of the parameter $D$ being negative
and close to that in (52). We emphasize that the energy of $T_{0}=386.6\,$%
MeV, at which the calculated phase shift changes sign and which is in good
agreement with the experimental value in (53), corresponds to the above
value of $D=-0.2147\,$fm$^{2}$. Calculated values of the pole parameters $D$
and $T_{0}$, and of the accuracy $\Delta $ are given in Table 2 for
different energy intervals. As can be seen from Table 2, the pole parameter $%
D$ is a quantity that admits an unambiguous determination yielding results
that are rather weakly dependent on the interval of fitting. In fact, any
fit to the phase shift leads, even at low energies, to negative values of
the parameter $D$ that are close to the value in (52). Thus, we see that, in
just the same way as the parameters $a_{t}$, $r_{t}$, and $v_{2}$, $D$ is a
low-energy parameter that can be determined reliably and to a high precision.

\vspace{1pt}

The pole formula (29) for the function $k\cot \delta $ coincides in form
with the well-known empirical Cini-Fubini-Stanghellini (CFS) formula \lbrack
19\rbrack , which was used to describe nucleon-nucleon scattering \lbrack
19, 31--35\rbrack\ and which is generally written as 
\begin{equation}
k\cot \delta =-\frac{1}{a_{t}}+\frac{1}{2}r_{t}k^{2}-\frac{pk^{4}}{1+qk^{2}}%
\,.  \tag{56}
\end{equation}
A distinctive feature of the Cini-Fubini-Stanghellini approach is that the
parameters $p\equiv -v_{2}$ and $q\equiv D$ in formula (56) are not
independent --- they are determined on the basis of the one-pion-exchange
(OPE) theory of nucleon-nucleon interaction. Thus, the quantities $p$ and $q$
in (56) are rather complicated functions of the scattering length $a_{t}$,
the effective range $r_{t}$, the pion mass $m_{\pi }$, and the pion-nucleon
coupling constant $G^{2}$. According to one-pion-exchange theory, the
expressions for the parameters $q$ and $p$ can be written in the form
\lbrack 32, 33\rbrack 
\begin{equation}
q=\rule[5pt]{4.6pt}{0.5pt}\hspace{-0.5em}\lambda _{\pi }^{2}\,\frac{%
2-f^{2}M\left( 3/\sqrt{2}-4\rule[5pt]{4.6pt}{0.5pt}\hspace{-0.5em}\lambda
_{\pi }/\hspace{-0.5em}\,\,a_{t}-r_{t}/2\rule[5pt]{4.6pt}{0.5pt}\hspace{%
-0.5em}\lambda _{\pi }\right) }{1-f^{2}M\left( 1/2\sqrt{2}-%
\rule[5pt]{4.6pt}{0.5pt}\hspace{-0.5em}\lambda _{\pi }/\hspace{-0.5em}%
\,\,a_{t}\right) }\,,  \tag{57}
\end{equation}
\begin{equation}
p=\left( \rule[5pt]{4.6pt}{0.5pt}\hspace{-0.5em}\lambda _{\pi
}^{2}-q/2\right) \left( 2\sqrt{2}\rule[5pt]{4.6pt}{0.5pt}\hspace{-0.5em}%
\lambda _{\pi }-r_{t}-4\rule[5pt]{4.6pt}{0.5pt}\hspace{-0.5em}\lambda _{\pi
}^{2}/a_{t}\right) \,,  \tag{58}
\end{equation}
where $M\equiv m_{N}/m_{\pi }$ is the ratio of the nucleon mass to the pion
mass, $\rule[5pt]{4.6pt}{0.5pt}\hspace{-0.5em}\lambda _{\pi }\equiv \hbar
/m_{\pi }c$ is the pion Compton wave length, and $f^{2}\equiv \left( m_{\pi
}/2m_{N}\right) ^{2}G^{2}$. For the low-energy scattering parameters $a_{t}$
and $r_{t}$, we will use, in the following, the values obtained above and
quoted in (49) and (50). On this basis, one can readily calculate the shape
parameter $v_{2}$ and the pole parameter $D$ within the
Cini-Fubini-Stanghellini approach. The results are 
\begin{equation}
v_{2}^{\text{CFS}}=-0.121\,\text{fm}^{3}\,,  \tag{59}
\end{equation}
\begin{equation}
D^{\text{CFS}}=3.777\,\text{fm}^{2}\,.  \tag{60}
\end{equation}

\vspace{1pt}

\vspace{1pt}It can be seen that the parameters $v_{2}$ and $D$ calculated
within the Cini-Fubini-Stanghellini approach differ considerably from the
``experimental'' values that are quoted in (51) and (52) and which are
determined quite reliably from present-day data on the triplet phase shift
\lbrack 39\rbrack . It should be noted that the distinction is not only
quantitative but also qualitative since the parameters $v_{2}$ and $D$ have
opposite signs within the Cini-Fubini-Stanghellini approach. The phase shift
is described poorly with the Cini-Fubini-Stanghellini parameters. The
explanation for so sharp a discrepancy with experimental data is likely to
be the following. It has been firmly established that one-pion exchange is
not the only mechanism and even is not the main mechanism of nucleon-nucleon
interaction --- the contribution of other mechanisms to nucleon-nucleon
interaction is much more significant. Therefore, it comes as no surprise
that the oversimplified one-pion-exchange scheme predicts erroneous values
for the parameters $v_{2}$ and $D$. For want of a theory that would make it
possible to calculate the low-energy parameters of nucleon-nucleon
interaction on the basis of a microscopic approach (QCD), one has to treat
them as adjustable parameters that are determined directly from experimental
data.

\vspace{1pt}

Along with the aforesaid, we note that the recent calculation of the shape
parameter $v_{2}$ within effective field theory (EFT) in \lbrack 40,
41\rbrack\ also yielded an incorrect sign of $v_{2}$: $v_{2}^{\text{EFT}%
}=-0.95\,$fm$^{3}$, although calculations within this theory generally led
to good agreement with experimental data for many computed features of the
nucleon-nucleon and neutron-deuteron systems (see references quoted in
\lbrack 40, 41\rbrack ). These discrepancies indicate that the shape
parameter $v_{2}$ and the higher order parameters $v_{n}$ are rather subtle
and sensitive characteristics of the nucleon-nucleon interaction. In \lbrack
40, 41\rbrack , it was also indicated that the contribution to the
nucleon-nucleon interaction from more ``short-range'' mechanisms than the
one-pion-exchange mechanism is of importance.

\vspace{1pt}\vspace{1pt}\vspace{1pt}

\begin{center}
\vspace{1pt}4. LOW-ENERGY PARAMETERS OF THE EFFECTIVE-RANGE EXPANSION
\end{center}

\vspace{1pt}

\vspace{1pt}Much attention has permanently been given to studying the
low-energy parameters of the effective-range expansion (11) \lbrack 12--15,
32--35, 37, 38, 40--50\rbrack . It should be noted that, while the
scattering length $a$ and the effective range $r_{0}$ can be determined
directly from experimental data to a fairly high degree of precision \lbrack
42--45\rbrack , the higher order parameters $v_{n}$ ($n=2,\,3,\,4,\,\ldots $%
) are less convenient for an experimental determination, their theoretical
calculation becoming more involved as the parameter order increases. At the
same time, the shape parameters $v_{n}$, along with the scattering length $a$
and the effective range $r_{0}$, are of particular importance for
constructing and comparing various realistic models of nucleon-nucleon
interaction and for describing nucleon-nucleon scattering. Moreover, the
shape parameters are of importance for investigating the physical properties
of the bound state in the two-nucleon system (deuteron) --- in particular,
their role in the useful expansion for the root-mean-square radius of the
deuteron was demonstrated in \lbrack 51--55\rbrack . The correlations
between the properties of the deuteron and the parameters of nucleon-nucleon
scattering that are determined by the expansion in (11) are also studied
\lbrack 51--56\rbrack . It should be noted that, in the past years, the
parameters $v_{n}$ were discussed and calculated in considering
nucleon-nucleon interaction within effective field theory \lbrack 40,
41\rbrack .

\vspace{1pt}

\vspace{1pt}In connection with the aforesaid, the determination of the
parameters of the effective-range expansion (11) is of great importance. An
explicit expansion of the function $k\cot \delta $ in a power series in $%
k^{2}$ follows directly from the pole representation (29): 
\begin{equation}
k\cot \delta =-\frac{1}{a_{t}}+\frac{1}{2}r_{t}k^{2}+\stackrel{\infty }{%
\stackunder{n=2}{\sum }}v_{2}\left( -D\right) ^{n-2}k^{2n}\,.  \tag{61}
\end{equation}
Therefore, all of the shape parameters $v_{n}$ can be also obtained
explicitly for this case. We have 
\begin{equation}
v_{n}=\left( -1\right) ^{n}v_{2}D^{n-2}\,,\,\,\,n\geqslant 3\,.  \tag{62}
\end{equation}
Thus, we see that, in the pole approximation, all parameters $v_{n}$ ($%
n=3,\,4,\,5,\,\ldots $) of the effective-range expansion are determined
explicitly in terms of the shape parameter $v_{2}$ and the pole parameter $D$
according to the simple formula (62) and can easily be calculated to any
order $n$.

\vspace{1pt}

It should be noted that\ the dimensionless shape parameters $p_{n}$\ are
also used along with the dimensional shape parameters\ $v_{n}$\ \lbrack see
Eq. (30)\rbrack . We define the dimensionless shape parameters $p_{n}$\
according to the relation 
\begin{equation}
v_{n}=p_{n}r_{t}^{2n-1}\,.  \tag{63}
\end{equation}
It can easily be seen that the parameters $p_{n}$\ are coefficients of the
power expansion of the function $kr_{t}\cot \delta _{t}$ with respect to\
the dimensionless parameter $x\equiv kr_{t}$. The dimensionless shape
parameters $p_{n}$ are convenient for qualitative analysis of the dependence
of the phase shift on energy. In our case, the parameter $D$ is negative by
virtue of (52). Since we also have in our case $p_{2}\equiv P_{t}=0.0304>0$
and $\left| D/r_{t}^{2}\right| <1$, one can easily see that all of the
parameters $p_{n}$, defined according to Eqs. (62) and (63), are positive
and decrease in absolute value; that is, 
\begin{equation}
p_{2}>p_{3}>\ldots >p_{n}>p_{n+1}>\ldots \,>0\,.  \tag{64}
\end{equation}

\vspace{1pt}

In Table 3, the low-energy scattering parameters $a_{t}$, $r_{t}$, and $%
v_{2} $ are presented along with the parameters $v_{3}$, $v_{4}$\ and $v_{5}$
calculated by formula (63) at the experimental value of $D=-0.225526\,$fm$%
^{2}$ and the shape-parameter values of $v_{2}=0.163\,$fm$^{3}$
(approximation P1) and $v_{2}=0.168\,$fm$^{3}$ (approximation P2). For the
sake of comparison, the parameters of the expansion in (11) that were found
in \lbrack 50\rbrack\ by using the partial-wave-analysis data on
nucleon-nucleon scattering that were obtained by the Nijmegen group \lbrack
57\rbrack\ (Nijm version) are also given in Table 3. We note that, at the
present time, the partial-wave-analysis data obtained by the VPI/GWU group
of Arndt et al. \lbrack 39\rbrack\ and the Nijmegen group \lbrack 57\rbrack\
are the most accurate and frequently used data on the phase shifts for
nucleon-nucleon scattering. It can be seen from Table 3 that the values
found here for the shape parameters $v_{n}$ ($n=3,\,4,\,5$) on the basis of
the pole approximation of the effective-range function are rather stable
with respect to the variations in the shape parameter $v_{2}$. It can easily
be shown that the parameters $v_{n}$ are also rather weakly sensitive to the
variation in the pole parameter $D$. Thus, formula (62) ensures a stable
determination of the parameters $v_{n}$.

\vspace{1pt}

\vspace{1pt} The following important comment concerning the applicability
and accuracy of the simple one-pole approximation (29) for determining the
parameters $v_{n}$ is in order. Since the smooth interpolation curve
specified by (29) provides an excellent description of experimental data of
Arndt et al. on the phase shift \lbrack 39\rbrack\ over a wide energy range
(in particular, to a precision not poorer than $0.01^{\circ }$ for energies
up to $T_{\text{lab}}=40\,$MeV), this curve determines the parameters $v_{n}$
with the degree of reliability and stability as high as that to which they
can in principle be determined by present-day experimental data on the phase
shift, the error in these data being $\gtrsim 0.01^{\circ }$. This means
that, if a different interpolation curve also described well the
experimental phase shift and, for some parameters $v_{n}$, gave values
considerably differing from those obtained here, one could conclude that
these parameters $v_{n}$ are not determined by experimental data. As a
matter of fact, the calculations show that only the parameters to $v_{4}$
inclusively are determined more or less reliably.

\vspace{1pt}

Table 3 shows that the results for the low-energy scattering parameters,
obtained here for the partial-wave-analysis data presented by Arndt et al.
\lbrack 39\rbrack\ are radically different from the results of the
calculations performed in \lbrack 50\rbrack\ for the partial-wave-analysis
data of the Nijmegen group \lbrack 57\rbrack , the distinction being not
only quantitative but also qualitative --- that is, all shape parameters
calculated here on the basis of the phase shifts of Arndt et al. are
positive and decrease, while the analogous parameters calculated for the
phase shifts of the Nijmegen group increase in absolute value and include a
negative parameter ($v_{4}$). To be more exact, the latter observation
concerns the dimensionless shape parameters $p_{n}$, which can be easily
calculated according to Eq. (63) and which are given in Table 4.

\vspace{1pt}

A detailed investigation of the parameters of the effective-range expansion
(11) will be given in our future publications. Here, we restrict ourselves
to mentioning briefly the following: in order to reveal possible reasons
behind such a discrepancy between the results, we have studied the stability
of the calculation of the shape parameters $v_{n}$ to variations in the
triplet scattering length $a_{t}$. We have shown that their sensitivity to
variations in this quantity is extremely high. In particular, a change in
the scattering length $a_{t}$ as small as a few tenths of a percent can lead
to a severalfold change in the shape parameter $v_{2}$. This is in accord
with the comment in the preceding section that the shape parameter $v_{2}$,
as well as the higher order parameters $v_{n}$, is a rather subtle and
sensitive characteristic of nucleon-nucleon interaction. Thus, so sharp a
distinction between the shape parameters $v_{n}$ for the GWU and the
Nijmegen phase shifts is due to a significant difference in the scattering
length: the value of $a_{t}^{\text{GWU}}=5.403\,$fm, which was obtained in
the present study, differs from $a_{t}^{\text{Nijm}}=5.420\,$fm more than by 
$0.3\%$. This difference in the values of the scattering length leads to a
decrease in the shape parameter $v_{2}$ by a factor of 4 for the phase
shifts of the Nijmegen group in relation to the phase shifts of the GWU
group --- that is, from the value of $v_{2}^{\text{GWU}}=0.163\,$fm$^{3}$ to
the value of $v_{2}^{\text{Nijm}}=0.040\,$fm$^{3}$. The discrepancy between
the corresponding higher order parameters $v_{n}$ is still larger. At the
same time, Table 3 shows that the values of the effective range $r_{t}$ for
the GWU and the Nijmegen phase shifts are close to each other. It should
also be noted that our results on the shape parameter $v_{2}$ are close to
the results of some earlier studies. In particular, the value of the shape
parameter $v_{2}=0.137\,$fm$^{3}$ corresponding to the dimensionless shape
parameter $P_{t}=0.027$ for the Reid soft-core potential RSC \lbrack
58\rbrack\ agrees well with the value of $v_{2}=0.163\,$fm$^{3}$, which was
obtained in the present study.

\vspace{1pt}

\vspace{1pt}We note that the Nijmegen value of the triplet scattering length 
$a_{t}^{\text{Nijm}}=5.420\,$fm is rather close to the presently recommended
experimental value \lbrack 59\rbrack 
\begin{equation}
a_{t}^{\text{expt}}=5.424\,\text{fm}\,,  \tag{65}
\end{equation}
while the value calculated here on the basis of the GWU phase shifts \lbrack
39\rbrack , $a_{t}^{\text{GWU}}=5.403\,$fm, is close to some of the
experimental values obtained previously for the triplet scattering length
\lbrack 32, 42, 43\rbrack . It should be emphasized that, for the
experimental values of the triplet scattering length $a_{t}$, various
authors \lbrack 32--35, 38, 39, 42--46, 58--61\rbrack\ present values
changing within a broad range --- from $5.37$ \lbrack 38, 60\rbrack\ to $%
5.479\,$fm \lbrack 61\rbrack . We also note that the present-day
experimental value of the triplet scattering length in (65) leads to
exaggerated (in relation to experimental data) values of the
root-mean-square radius $r_{d}$ of the deuteron \lbrack 52, 53, 56\rbrack\
and the asymptotic normalization constant $A_{S}$ for it \lbrack 56\rbrack .
From all of the aforesaid, we can therefore draw the following important
conclusion: in order to remove the existing discrepancies between the
current experimental value of $a_{t}$ in (65), on one hand, and the
present-day values of the quantities $r_{d}$ and $A_{S}$ for the deuteron
and the partial-wave-analysis data of the GWU group, which lead to the value
of $a_{t}^{\text{GWU}}=5.403\,$fm, on the other hand, it is of paramount
importance to refine the experimental value of the triplet scattering length 
$a_{t}$. The value of the triplet scattering length $a_{t}$ is also of
particular importance since it is often used as one of the input values in
fitting the parameters of nucleon-nucleon potentials. We note that the value
of $a_{t}^{\text{GWU}}=5.403\,$fm, which corresponds to the partial-wave
analysis performed by Arndt et al., is in perfect agreement with the
experimental values of the quantities $r_{d}$ and $A_{S}$ for the deuteron
\lbrack 56\rbrack .

\vspace{1pt}

In a wide energy range \lbrack approximately to the energy corresponding to
the pole of the function $k\cot \delta $ (53)\rbrack , the experimental
triplet phase shift \lbrack 39\rbrack\ is well described by the
effective-range expansion (11) with the low-energy scattering parameters $%
a_{t},$ $r_{t},$ $v_{2},$ $v_{3},\ldots $ set to the values determined here.
This is a direct corollary of the fact that, for the phase shift borrowed
from \lbrack 39\rbrack , all of the dimensionless shape parameters $p_{n}$
appear to be decreasing in absolute value. For example, a fifth-degree
polynomial in $k^{2}$ describes the experimental phase shift in the energy
range $T_{\text{lab}}\leqslant 200\,$MeV to within about $1^{\circ }$ and in
the energy range $T_{\text{lab}}\leqslant 50\,$MeV to within about $%
0.02^{\circ }$. The description of the function $k\cot \delta $ by
various-degree polynomials corresponding to various polynomial
approximations of the effective-range expansion (11) are shown in Fig. 3. We
see that a successive increase in the degree of a polynomial leads to an
improved description of the function $k\cot \delta $ (and, accordingly, of
the phase shift) and to the extension of the interval where this description
is valid. Thus, the use of the effective-range expansion with a small number
of terms provides a good description of the phase shift over a wide energy
range ($0-250\,$MeV) if the low-energy parameters are set to the values
found here. This indicates that, in contrast to statements advocated in some
articles (see, for example, \lbrack 50\rbrack ), the effective-range
expansion is highly accurate and very useful in describing nucleon-nucleon
scattering.

\vspace{1pt}

The potential of the effective-range expansion (11) is directly related to
the radius of its convergence. The function $k\cot \delta $ considered in
the complex plane of $k^{2}$ is an analytic function of $k^{2}$ within some
region near the origin of coordinates; therefore, it can be expanded in a
Taylor-Maclaurin series (11) in powers of $k^{2}$ in the vicinity of the
point $k^{2}=0$. Thus, a particular importance of the function $k\cot \delta 
$ is associated with its analyticity near the point $k^{2}=0$. Noyes and
Wong \lbrack 62\rbrack\ showed that the one-pion-exchange model for
nucleon-nucleon interaction leads to the appearance of a cut in the
scattering amplitude on the negative axis of energy for $k^{2}\leqslant
-m_{\pi }^{2}c^{2}/4\hbar ^{2}$; therefore, the radius of convergence of the
series in (11) is extremely small in the one-pion-exchange model, $%
T_{0}=m_{\pi }^{2}/2m_{N}=9.7\,$MeV. This contradicts the fact that the
experimental phase shift \lbrack 39\rbrack\ is well described by the pole
formula (29) with the parameters set to the values in (49)--(52), since, for
the radius of convergence of the series in (11), formula (29) gives the
value of $T_{0}=368.026\,$MeV, which we chose on the basis of the
``experimental'' condition $\delta \left( T_{0}\right) =0$ (the convergence
circle is determined by the condition $\left| k^{2}\right| <1/\left|
D\right| $). In connection with this contradiction, we recall the comments
at the end of the preceding section that concern the limited applicability
of the one-pion-exchange mechanism. As was indicated by Noyes himself
\lbrack 33\rbrack , there is no a priori method for estimating the accuracy
or the region of applicability of the phenomenological expansion (11), so
that this question must be solved on the basis of experimental data.

\vspace{1pt}

Numerical values of the dimensionless shape parameters $p_{n}$ provide
direct information about the radius of convergence of the series in (11). In
the case under consideration, the increase in the parameters $p_{n}$ in
absolute value indicates, as is the case for the triplet $np$ phase shift of
the Nijmegen group, that the effective-range expansion (11) has a small
radius of convergence, so that it is not very useful \lbrack 50\rbrack . On
the contrary, the decrease in the shape parameters $p_{n}$ for the triplet $%
np$ phase shift of the GWU group suggests that the radius of convergence of
the series in (11) is large and that this expansion is very useful in this
case. This is corroborated by a high quality of the description of the phase
shift in the case where the function $k\cot \delta $ is approximated by
polynomials of relatively low degrees. Thus, the present-day data from the
partial-wave analyses performed by the two main groups in \lbrack 39\rbrack\
and \lbrack 57\rbrack\ do not agree with each other both in what is
concerned with the low-energy scattering parameters calculated on their
basis and in what is concerned with the applicability of the effective-range
expansion to them. At the same time, it is important to note that the
numerical values obtained for the phase shifts by the GWU and the Nijmegen
group are rather close to each other. This indicates once again that the
low-energy scattering parameters are subtle and sensitive characteristics of
nucleon-nucleon scattering.

\vspace{1pt}

\vspace{1pt}\vspace{1pt}

\begin{center}
\vspace{1pt}5. DESCRIPTION OF THE PROPERTIES OF THE DEUTERON
\end{center}

\vspace{1pt}\vspace{1pt}\vspace{1pt}

\vspace{1pt}Following the same line of reasoning as in the case of
constructing the effective-range expansion (11), one can write an expansion
of the function $k\cot \delta $ in a power series at the point $%
k^{2}=-\alpha ^{2}$ ---\ that is, at the energy equal to the deuteron
binding energy $\varepsilon _{d}=\hbar ^{2}\alpha ^{2}/m_{N\text{ }}$. This
expansion has the form 
\begin{equation}
k\cot \delta =-\alpha +\frac{1}{2}\rho _{d}\left( k^{2}+\alpha ^{2}\right)
+w_{2}\left( k^{2}+\alpha ^{2}\right) ^{2}+\ldots \,,  \tag{66}
\end{equation}
where $\rho _{d}\equiv \rho \left( -\varepsilon _{d},-\varepsilon
_{d}\right) $ is the deuteron effective range corresponding to $S$-wave
interaction. The definition and the properties of the effective deuteron
range $\rho _{d}$ and of the function $\rho \left( E_{1},E_{2}\right) $ are
discussed in detail elsewhere \lbrack 38\rbrack . Using Eqs. (29) and (66),
we can easily establish that the quantities $\alpha $ and $\rho _{d}$ for
the deuteron are related to the parameters of the pole representation of the
effective-range function as 
\begin{equation}
\alpha =\frac{1}{a_{t}}+\frac{1}{2}r_{t}\alpha ^{2}-\frac{v_{2}\alpha ^{4}}{%
1-D\alpha ^{2}}\,,  \tag{67}
\end{equation}
\begin{equation}
\rho _{d}=\rho _{m}-\frac{2v_{2}\alpha ^{2}}{\left( 1-D\alpha ^{2}\right)
^{2}}\,,  \tag{68}
\end{equation}
where $\rho _{m}\equiv \rho \left( 0,-\varepsilon _{d}\right) $ is the
so-called mixed effective range given by 
\begin{equation}
\rho _{m}=\frac{2}{\alpha }\left( 1-\frac{1}{\alpha a_{t}}\right) \,. 
\tag{69}
\end{equation}
Using formulas (67) and (69), we can recast the expression for the mixed
effective range $\rho _{m}$ into the form 
\begin{equation}
\rho _{m}=r_{t}-\frac{2v_{2}\alpha ^{2}}{1-D\alpha ^{2}}\,.  \tag{70}
\end{equation}

\vspace{1pt}

From formulas (70) and (68), one can easily obtain expansions of the
standard scattering effective range $r_{t}$ and the deuteron effective range 
$\rho _{d}$ in power series in $\alpha ^{2}$. We have 
\begin{equation}
r_{t}=\rho _{m}-2\stackrel{\infty }{\stackunder{n=1}{\sum }}\left( -1\right)
^{n}v_{n+1}\alpha ^{2n}\,,  \tag{71}
\end{equation}
\begin{equation}
\rho _{d}=\rho _{m}+2\stackrel{\infty }{\stackunder{n=1}{\sum }}\left(
-1\right) ^{n}n\,v_{n+1}\alpha ^{2n}\,,  \tag{72}
\end{equation}
where the parameters $v_{n}$ are given by (62). The dimensionless asymptotic
normalization factor $C_{d}$ for the deuteron is expressed in terms of the
deuteron effective range $\rho _{d}$ as 
\begin{equation}
C_{d}^{2}=\left( 1-\alpha \rho _{d}\right) ^{-1}\,.\,  \tag{73}
\end{equation}
The constant $C_{d}$ and the asymptotic normalization factor $A_{S}$, which
is widely used in the literature \lbrack 47, 63\rbrack , are related by the
equation 
\begin{equation}
A_{S}^{2}=2\alpha C_{d}^{2}\,.\,  \tag{74}
\end{equation}

\vspace{1pt}

\vspace{1pt}For various approximations of the effective-range function, we
have calculated the following parameters characterizing the deuteron: the
binding energy $\varepsilon _{d}$, the effective range $\rho _{d}$, and the
asymptotic normalization factors $C_{d}$ and $A_{S}$. In Table 5, the
results of these calculations are given along with their experimental
counterparts from \lbrack 64, 65\rbrack . We note that the deuteron binding
energy was calculated by using a relativistic formula, which is more
accurate than (21). It can be seen from Table 5 that the features of the
deuteron that were calculated in the pole approximation (29) with the
parameter values from (49)--(52) are in good agreement with their
experimental values. The results obtained in the shape-parameter
approximation (SP) differ insignificantly from their counterparts in the
pole approximation, while the results of the calculations that take into
account the cubic term in energy in the effective-range expansion are nearly
coincident with those in the pole approximation. It can be seen from Table 5
that the convergence of the calculated features of the deuteron versus the
number of terms that are taken into account in the effective-range expansion
is very fast --- the shape-parameter approximation yields a highly precise
result, the inclusion of higher order terms in energy introduces virtually
no changes in this result. This is a consequence of the fact that the
dimensionless shape parameters $p_{n}$ are decreasing quantities in this
case. It should be emphasized that the features of the deuteron were
calculated on the basis of the scattering-parameter values in (49)--(52),
which correspond to the phase shifts of Arndt et al., and this means that we
have very good agreement between the experimental data of the GWU group on
scattering \lbrack 39\rbrack\ and the experimental data on the bound state
(deuteron) from \lbrack 64, 65\rbrack .

\vspace{1pt}

The values found for the features of the deuteron in the effective-range
approximation (ER) are somewhat exaggerated in relation to their
experimental counterparts and the results obtained in the pole
approximation. This is due primarily to an insufficiently accurate
determination of the deuteron binding energy in the effective-range
approximation with the low-energy parameters $a_{t}=5.403\,$fm and $%
r_{t}=1.7494\,$fm. If one uses the experimental values for the deuteron
binding energy and for the triplet scattering length ($\varepsilon
_{d}=2.224589\,$MeV and $a_{t}=5.403\,$fm, respectively), the
effective-range approximation yields the following values for the deuteron
effective range $\rho _{d}$ and for the asymptotic normalization factor $%
A_{S}$: $\rho _{d}=1.7331\,$fm and $A_{S}=0.8795\,$fm$^{-1/2}$, these
results being in good agreement with the corresponding experimental values
and with the results of the calculations in the pole approximation. This is
in accord with the results obtained in \lbrack 56\rbrack , where it was
established that the asymptotic normalization factor $A_{S}$ depends only
slightly on the model of nucleon-nucleon interaction; at the experimental
value of the deuteron binding energy $\varepsilon _{d}$, $99.7\%$ of it is
determined by the triplet scattering length $a_{t}$.

\vspace{1pt}

\ \vspace{1pt}

\begin{center}
\vspace{1pt}6. PARAMETERS OF THE BARGMANN $S$ MATRIX
\end{center}

\vspace{1pt}\vspace{1pt}\vspace{1pt}

The parameters $\alpha $, $\beta $, $\lambda $, and $\mu $ of the Bargmann $%
S $ matrix (20), which corresponds to the presence of two physical states in
the system, are unambiguously related to the parameters of the pole
approximation of the effective-range function (29) and, as can easily be
seen, are the roots of the fourth-degree algebraic equation 
\begin{equation}
v_{2}x^{4}-\left( 1-Dx^{2}\right) \left( \frac{1}{a_{t}}+\frac{1}{2}%
r_{t}x^{2}-x\right) =0\,.  \tag{75}
\end{equation}
Solving Eq. (75) with the parameters set to the values in (49)--(52), we
obtain the following values for the parameters of the Bargmann $S$ matrix: 
\begin{equation}
\alpha =0.2315\,\text{fm}^{-1},\,\beta =1.2293\,\text{fm}^{-1},\,\lambda
=2.5603+i3.5248\,\text{fm}^{-1},\,\mu =2.5603-i3.5248\,\text{fm}^{-1}. 
\tag{76}
\end{equation}

\vspace{1pt}

In the case being considered, the two-nucleon system has two bound states.
In accordance with the phenomenology of nodes that employs potentials
involving forbidden states \lbrack 16--18\rbrack , the lowest, deeply lying,
state characterized by the energy 
\begin{equation}
\varepsilon _{0}=63.7555\,\text{MeV}\,,  \tag{77}
\end{equation}
is unobservable, while the excited state of binding energy 
\begin{equation}
\varepsilon _{d}=2.2237\,\text{MeV}  \tag{78}
\end{equation}
corresponds to the deuteron.

\vspace{1pt}

\vspace{1pt}Thus, the use of the pole approximation in describing triplet
neutron-proton scattering automatically leads to considering deep potentials
involving forbidden states.

\vspace{1pt}

In the shape-parameter approximation (44) ($D=0$), we have the following
values for the parameters of the Bargmann $S$ matrix (20): 
\begin{equation}
\alpha =0.2315\,\text{fm}^{-1},\,\beta =-2.7787\,\text{fm}^{-1},\,\lambda
=1.2736+i0.3784\,\text{fm}^{-1},\,\mu =1.2736-i0.3784\,\text{fm}^{-1}. 
\tag{79}
\end{equation}
In this case, the second state is a virtual state at the energy 
\begin{equation}
\varepsilon _{v}=353.4732\,\text{MeV}\,,  \tag{80}
\end{equation}
while the ground state, whose binding energy is 
\begin{equation}
\varepsilon _{d}=2.2236\,\text{MeV}  \tag{81}
\end{equation}
\lbrack in fact, it coincides with that in (78)\rbrack , corresponds to the
deuteron. As might have been expected, the parameters in (79) satisfy
relation (43).

\vspace{1pt}\vspace{1pt}

\begin{center}
\vspace{1pt}7. CONCLUSION
\end{center}

\vspace{1pt}\vspace{1pt}\vspace{1pt}

\vspace{1pt}Our basic results and conclusions can be formulated as follows.
Relying on the Bargmann representation of the $S$ matrix, we have formulated
the pole approximation for the effective-range function $k\cot \delta $. At
specific values of the $S$-matrix parameters, the effective-range
approximation and the shape-parameter approximation immediately follow from
this approximation. The pole approximation of the function $k\cot \delta $
is optimal for describing nucleon-nucleon scattering and involves a few
parameters. The parameters of this approximation have a clear physical
meaning. They are related to the parameters of the Bargmann $S$ matrix by
simple equations. It has been shown that the pole approximation deduced from
the Bargmann representation of the $S$ matrix is a direct generalization of
the effective-range approximation to the case where there are two physical
states in the system. The presence of the pole term makes it possible to
improve significantly, by using only two additional parameters, the
description of the phase shift and to expand sizably the applicability range
of this description. In the shape-parameter approximation corresponding to
the Bargmann representation of the $S$ matrix, one can obtain an important
constraint on the dimensionless shape parameter, $P>0$ ---\ that is, the
shape parameter is positive.

\vspace{1pt}

By using a least squares polynomial fit to the function $k\cot \delta $ at
low energies, we have obtained, on the basis of the analysis of the latest
experimental data on phase shifts that was performed by the VPI/GWU group,
the triplet low-energy parameters of neutron-proton scattering: $%
a_{t}=5.4030\,$fm, $r_{t}=1.7494\,$fm, and $v_{2}=0.163\,$fm$^{3}$. With
these values of $a_{t}$, $r_{t}$, and $v_{2}$ and the pole parameter $D$,
the pole approximation of the function $k\cot \delta $ provides an excellent
description of the triplet phase shift for neutron-proton scattering over a
wide energy range ($T_{\text{lab}}\lesssim 1000\,$MeV), the description in
the low-energy region also being improved considerably.

\vspace{1pt}

For the experimental phase shifts of the VPI/GWU group, the values that we
have obtained on the basis of the pole approximation for the triplet shape
parameters of the effective-range expansion are positive and decrease with
increasing $n$. On the basis of the effective-range expansion with the
values found for the low-energy scattering parameters $a_{t},$ $r_{t},$ $%
v_{2},$ $v_{3},\ldots \,$, the phase shift is described well over a wide
energy range extending approximately to the energy at which the phase shift
changes sign, this being a direct consequence of a decrease in the
dimensionless shape parameters with increasing $n$. This circumstance is
indicative of a high precision of the effective-range expansion and its high
potential for describing experimental data on nucleon-nucleon scattering, in
contrast to the statements of some authors (see, for example, \lbrack
50\rbrack ).\ 

\vspace{1pt}

\vspace{1pt}The results obtained here for the shape parameters by using the
data of the partial-wave analysis performed by the GWU group differ
drastically from the results of the calculations in \lbrack 50\rbrack\ for
the data of the partial-wave analysis performed by the Nijmegen group, this
distinction being not only quantitative but also qualitative --- that is,
all of the dimensionless shape parameters calculated here by using the phase
shifts of the GWU group are positive and decrease, while the analogous
parameters calculated for the Nijmegen phase shifts increase in absolute
value and include a negative parameter ($p_{4}$). In our opinion, so sharp a
discrepancy between the shape parameters for the GWU and the Nijmegen phase
shifts is due to quite a significant difference in the scattering length:
the value $a_{t}^{\text{GWU}}=5.403\,$fm obtained here differs from $a_{t}^{%
\text{Nijm}}=5.420\,$fm more than by $0.3\%$. This difference in the
scattering length leads to the decrease in the shape parameter $v_{2}$ by a
factor of 4 for the Nijmegen phase shifts in relation to the phase shifts
obtained by the GWU group --- that is, from the value $v_{2}^{\text{GWU}%
}=0.163\,$fm$^{3}$ to the value of $v_{2}^{\text{Nijm}}=0.040\,$fm$^{3}$
---\ the discrepancy between the corresponding values of the higher order
parameters $v_{n}$ being still greater. This confirms that the shape
parameter $v_{2}$, as well as the higher order parameters $v_{n}$, is a
rather subtle and sensitive characteristic of nucleon-nucleon interaction.
The aforesaid leads to the important conclusion that an experimental
refinement of the triplet scattering length $a_{t}$ is of paramount
importance since it is necessary to remove the existing discrepancies
between the current experimental value of $a_{t}^{\text{expt}}=5.424\,$fm,
on one hand, and the present-day values of $r_{d}$ and $A_{S}$ for the
deuteron \lbrack 56\rbrack\ and the present-day data of the partial-wave
analysis performed by the GWU group, which lead to the value of $a_{t}^{%
\text{GWU}}=5.403\,$fm, on the other hand.

\vspace{1pt}

For various approximations of the effective-range function, we have
calculated the main features of the deuteron --- the binding energy $%
\varepsilon _{d}$, the effective range $\rho _{d}$, and the asymptotic
normalization factors $C_{d}$ and $A_{S}$. The results obtained for them in
the pole approximation agree very well with their experimental counterparts.
The results in the shape-parameter approximation differ insignificantly from
those in the pole approximation. We have found that the convergence of the
calculated features of the deuteron versus the number of terms that are
taken into account in the effective-range expansion is very fast --- the
shape-parameter approximation gives a nearly precise result, which undergoes
virtually no changes upon taking into account higher order terms in energy.
Thus, we can state that, on the basis of the Bargmann representation of the $%
S$ matrix, a good unified description has been obtained for the bound state
of the two-nucleon system and the triplet phase shift for neutron-proton
scattering up to energies of about $1000\,$MeV.

\vspace{1pt}\vspace{1pt}

\begin{center}
REFERENCES
\end{center}

\begin{enumerate}
\item  A. G. Sitenko, \textit{Scattering Theory} (Springer-Verlag, Berlin,
1991).

\item  W. Heisenberg, Z. Phys. \textbf{120, }513 (1943).

\item  R. Jost, Helv. Phys. Acta \textbf{20}, 256 (1947).

\item  R. G. Newton, \textit{Scattering Theory of Waves and Particles}, 2nd
ed. (Springer-Verlag, New York, 1982).

\item  V. A. Babenko and N. M. Petrov, Phys. At. Nucl. \textbf{63}, 1709
(2000).

\item  V. V. Malyarov and M. N. Popushoi, Sov. J. Nucl. Phys. \textbf{18},
586 (1973).

\item  B. N. Zakhar'ev, V. N. Pivovarchik, E. B. Plekhanov, \textit{et al.},
Sov. J. Part. Nucl. \textbf{13}, 535 (1982).

\item  B. N. Zakhar'ev, P. Yu. Nikishev, and E. B. Plekhanov, Sov. J. Nucl.
Phys. \textbf{38}, 54 (1983).

\item  V. Bargmann, Rev. Mod. Phys. \textbf{21}, 488\ (1949).

\item  W. T. H. van Oers and J. D. Seagrave, Phys. Lett. B\textbf{\ 24},
562\ (1967).

\item  G. A. Baker, Jr. and P. Graves-Morris, \textit{Pad\'{e} Approximants}
(Addison-Wesley, London, 1981).

\item  F. Lambert, O. Corbella, and Z. D. Thome, Nucl. Phys. B\textbf{\ 90, }%
267 (1975).

\item  K. Hartt, Phys. Rev. C\textbf{\ 22}, 1377\ (1980).

\item  K. Hartt, Phys. Rev. C\textbf{\ 23}, 2399\ (1981).

\item  K. Hartt, Bull. Am. Phys. Soc. \textbf{23}, 630\ (1978).

\item  V. G. Neudatchin, I. T. Obukhovsky, V. I. Kukulin, \textit{et al.},
Phys. Rev. C\textbf{\ 11}, 128\ (1975).

\item  V. G. Neudatchin, I. T. Obukhovsky, and Yu. F. Smirnov, Sov. J. Part.
Nucl. \textbf{15}, 519 (1984).

\item  V. I. Kukulin, V. M. Krasnopolskij, V. N. Pomerantsev, \textit{et al.}%
, Sov. J. Nucl. Phys. \textbf{43}, 355 (1986).

\item  M. Cini, S. Fubini, and A. Stanghellini, Phys. Rev. \textbf{114},
1633\ (1959).

\item  R. S. Christian and J. L. Gammel, Phys. Rev. \textbf{91}, 100\ (1953).

\item  L. M. Delves, Phys. Rev. \textbf{118}, 1318\ (1960).

\item  A. S. Reiner, Phys. Lett. B\textbf{\ 28}, 387\ (1969).

\item  J. S. Whiting and M. G. Fuda, Phys. Rev. C\textbf{\ 14}, 18\ (1976).

\item  V. E. Kuz'michev, Ukr. Fiz. Zh. \textbf{23}, 188 (1978) \lbrack in
Russian\rbrack .

\item  B. A. Girard and M. G. Fuda, Phys. Rev. C\textbf{\ 19}, 579\ (1979).

\item  S. A. Adhikari and J. R. A. Torreao, Phys. Lett. B\textbf{\ 132},
257\ (1983).

\item  I. V. Simenog, A. I. Sitnichenko, and D. V. Shapoval, Sov. J. Nucl.
Phys. \textbf{45}, 37 (1987).

\item  C. R. Chen, G. L. Payne, J. L. Friar, and B. F. Gibson, Phys. Rev. C%
\textbf{\ 39}, 1261\ (1989).

\item  Yu. V. Orlov and L. I. Nikitina, Phys. At. Nucl. \textbf{61}, 750
(1998).

\item  Yu. V. Orlov, Yu. P. Orevkov, and L. I. Nikitina, Phys. At. Nucl. 
\textbf{65}, 371 (2002).

\item  D. Y. Wong and H. P. Noyes, Phys. Rev. \textbf{126}, 1866\ (1962).

\item  H. P. Noyes, Phys. Rev. \textbf{130}, 2025\ (1963).

\item  H. P. Noyes, Ann. Rev. Nucl. Sci. \textbf{22}, 465 (1972).

\item  L. Mathelitsch and B. J. VerWest, Phys. Rev. C\textbf{\ 29}, 739\
(1984).

\item  J. R. Bergervoet, P. C. van Campen, W. A. van der Sanden, and J. J.
de Swart, Phys. Rev. C\textbf{\ 38}, 15\ (1988).

\item  V. M. Galitsky, B. M. Karnakov, and V. I. Kogan, \textit{Problems in
Quantum Mechanics} (Nauka, Moscow, 1992) \lbrack in Russian\rbrack .

\item  J. M. Blatt and J. D. Jackson, Phys. Rev. \textbf{76}, 18\ (1949).

\item  L. Hulth\'{e}n and M. Sugawara, in \textit{Handbuch der Physik}, Ed.
by S. Fl\"{u}gge (Springer-Verlag, New York, Berlin, 1957), p. 1.

\item  R. A. Arndt, W. J. Briscoe, I. I. Strakovsky, and R. L. Workman, 
\textit{Partial-Wave Analysis Facility\ SAID}, The George Washington
University \lbrack http://gwdac.phys.gwu.edu\rbrack ; Phys. Rev. C \textbf{62%
}, 034005 (2000).

\item  T. D. Cohen and J. M. Hansen, Phys. Rev. C\textbf{\ 59}, 13\ (1999).

\item  T. D. Cohen and J. M. Hansen, Phys. Rev. C\textbf{\ 59}, 3047\ (1999).

\item  R. Wilson, \textit{The Nucleon-Nucleon Interaction} (Interscience,
New York, 1963).

\item  T. L. Houk and R. Wilson, Rev. Mod. Phys. \textbf{39}, 546\ (1967).

\item  T. L. Houk, Phys. Rev. C\textbf{\ 3}, 1886\ (1971).

\item  E. L. Lomon and R. Wilson, Phys. Rev. C\textbf{\ 9}, 1329\ (1974).

\item  W. Dilg, Phys. Rev. C\textbf{\ 11}, 103\ (1975).

\item  M. W. Kermode, A. McKerrell, J. P. McTavish, and L. J. Allen, Z.
Phys. A\textbf{\ 303, }167 (1981).

\item  C. W. Wong, Nucl. Phys. A\textbf{\ 536, }269 (1992).

\item  W. van Dijk, M. W. Kermode, and D.-C. Zheng, Phys. Rev. C\textbf{\ 47}%
, 1898\ (1993).

\item  J. J. de Swart, C. P. F. Terheggen, and V. G. J. Stoks, \textit{%
Invited talk at the 3rd International Symposium ``Dubna Deuteron 95'',
Dubna, Russia, July 4-7, 1995,} nucl-th/9509032.

\item  M. W. Kermode, S. A. Moszkowski, M. M. Mustafa, and W. van Dijk,
Phys. Rev. C\textbf{\ 43}, 416\ (1991).

\item  D. W. L. Sprung, in \textit{Proceedings of IX European Conference on
Few-Body Problems in Physics, Tbilisi, Georgia, USSR, Aug. 1984,} (World
Sci. Press, Singapore; Philadelphia, 1984), p. 234.

\item  S. Klarsfeld, J. Martorell, J. A. Oteo \textit{et al.}, Nucl. Phys. A%
\textbf{\ 456}, 373\ (1986).

\item  D. W. L. Sprung, H. Wu, and J. Martorell, Phys. Rev. C\textbf{\ 42},
863\ (1990).

\item  R. K. Bhaduri, W. Leidemann, G. Orlandini, and E. L. Tomusiak, Phys.
Rev. C\textbf{\ 42}, 1867\ (1990).

\item  V. A. Babenko and N. M. Petrov, Phys. At. Nucl. \textbf{66}, 1319
(2003).

\item  Nijmegen NN-Online program \lbrack http://nn-online.sci.kun.nl\rbrack
; V. G. J. Stoks, R. A. M. Klomp, M. C. M. Rentmeester, and J. J. de Swart,
Phys. Rev. C\textbf{\ 48}, 792\ (1993).

\item  R. V. Reid, Jr., Ann. Phys. \textbf{50}, 411\ (1968).

\item  O. Dumbrajs, R. Koch, H. Pilkuhn, \textit{et al.}, Nucl. Phys. B%
\textbf{\ 216}, 277\ (1983).

\item  G. L. Squires and A. T. Stewart, Proc. Roy. Soc. A\textbf{\ 230}, 19\
(1955).

\item  H. Kaiser, H. Rauch, G. Badurek, \textit{et al.}, Z. Phys. A\textbf{\
291}, 231 (1979).

\item  H. P. Noyes and D. Y. Wong, Phys. Rev. Lett. \textbf{3}, 191\ (1959).

\item  T. E. O. Ericson, Nucl. Phys. A\textbf{\ 416, }281 (1984).

\item  G. L. Greene, E. G. Kessler Jr., R. D. Deslattes, and H. Boerner,
Phys. Rev. Lett. \textbf{56}, 819\ (1986).

\item  I. Borb\'{e}ly, W. Gr\"{u}ebler, V. K\"{o}nig, \textit{et al.}, Phys.
Lett. B\textbf{\ 160}, 17\ (1985).
\end{enumerate}

\newpage \noindent \textbf{Table 1.} Triplet phase shift for $np$ scattering
as a function of the laboratory energy $T_{\text{lab}}$ according to
calculations within the effective-range (ER) approximation, the
shape-parameter (SP) approximation, and the pole (P1 and P2) approximations
(for the shape-parameter values $v_{2}=0.163\,$fm$^{3}$ and $v_{2}=0.168\,$fm%
$^{3}$, respectively).

\begin{center}
\vspace{1pt} 
\begin{tabular}{|c|ccccc|}
\hline
$T_{\text{lab}}\,$, &  & Phase & shift & $\delta _{t}\,$,$\,\,\,\,\ $deg & 
\\ \cline{2-6}
MeV & ER & \multicolumn{1}{|c}{SP} & \multicolumn{1}{|c}{P1} & 
\multicolumn{1}{|c}{P2} & \multicolumn{1}{|c|}{Experiment \lbrack 39\rbrack}
\\ \hline
$1$ & $147.84$ & \multicolumn{1}{|c}{$147.83$} & \multicolumn{1}{|c}{$147.83$%
} & \multicolumn{1}{|c}{$147.83$} & \multicolumn{1}{|c|}{$147.83$} \\ \hline
$5$ & $118.34$ & \multicolumn{1}{|c}{$118.23$} & \multicolumn{1}{|c}{$118.23$%
} & \multicolumn{1}{|c}{$118.23$} & \multicolumn{1}{|c|}{$118.23$} \\ \hline
$10$ & $102.93$ & \multicolumn{1}{|c}{$102.56$} & \multicolumn{1}{|c}{$%
102.55 $} & \multicolumn{1}{|c}{$102.54$} & \multicolumn{1}{|c|}{$102.55$}
\\ \hline
$25$ & $81.87$ & \multicolumn{1}{|c}{$80.36$} & \multicolumn{1}{|c}{$80.26$}
& \multicolumn{1}{|c}{$80.20$} & \multicolumn{1}{|c|}{$80.26$} \\ \hline
$40$ & $71.19$ & \multicolumn{1}{|c}{$68.44$} & \multicolumn{1}{|c}{$68.11$}
& \multicolumn{1}{|c}{$68.01$} & \multicolumn{1}{|c|}{$68.11$} \\ \hline
$50$ & $66.23$ & \multicolumn{1}{|c}{$62.68$} & \multicolumn{1}{|c}{$62.14$}
& \multicolumn{1}{|c}{$62.01$} & \multicolumn{1}{|c|}{$62.11$} \\ \hline
$75$ & $57.51$ & \multicolumn{1}{|c}{$52.16$} & \multicolumn{1}{|c}{$50.90$}
& \multicolumn{1}{|c}{$50.70$} & \multicolumn{1}{|c|}{$50.77$} \\ \hline
$100$ & $51.64$ & \multicolumn{1}{|c}{$44.80$} & \multicolumn{1}{|c}{$42.60$}
& \multicolumn{1}{|c}{$42.34$} & \multicolumn{1}{|c|}{$42.34$} \\ \hline
$125$ & $47.30$ & \multicolumn{1}{|c}{$39.25$} & \multicolumn{1}{|c}{$35.95$}
& \multicolumn{1}{|c}{$35.66$} & \multicolumn{1}{|c|}{$35.58$} \\ \hline
$150$ & $43.93$ & \multicolumn{1}{|c}{$34.89$} & \multicolumn{1}{|c}{$30.37$}
& \multicolumn{1}{|c}{$30.05$} & \multicolumn{1}{|c|}{$29.94$} \\ \hline
$175$ & $41.19$ & \multicolumn{1}{|c}{$31.35$} & \multicolumn{1}{|c}{$25.53$}
& \multicolumn{1}{|c}{$25.19$} & \multicolumn{1}{|c|}{$25.09$} \\ \hline
$200$ & $38.92$ & \multicolumn{1}{|c}{$28.42$} & \multicolumn{1}{|c}{$21.23$}
& \multicolumn{1}{|c}{$20.90$} & \multicolumn{1}{|c|}{$20.84$} \\ \hline
$225$ & $36.99$ & \multicolumn{1}{|c}{$25.95$} & \multicolumn{1}{|c}{$17.36$}
& \multicolumn{1}{|c}{$17.04$} & \multicolumn{1}{|c|}{$17.03$} \\ \hline
$250$ & $35.32$ & \multicolumn{1}{|c}{$23.84$} & \multicolumn{1}{|c}{$13.81$}
& \multicolumn{1}{|c}{$13.53$} & \multicolumn{1}{|c|}{$13.57$} \\ \hline
$275$ & $33.86$ & \multicolumn{1}{|c}{$22.01$} & \multicolumn{1}{|c}{$10.53$}
& \multicolumn{1}{|c}{$10.29$} & \multicolumn{1}{|c|}{$10.37$} \\ \hline
$300$ & $32.57$ & \multicolumn{1}{|c}{$20.42$} & \multicolumn{1}{|c}{$7.47$}
& \multicolumn{1}{|c}{$7.28$} & \multicolumn{1}{|c|}{$7.38$} \\ \hline
$325$ & $31.41$ & \multicolumn{1}{|c}{$19.02$} & \multicolumn{1}{|c}{$4.60$}
& \multicolumn{1}{|c}{$4.47$} & \multicolumn{1}{|c|}{$4.56$} \\ \hline
$350$ & $30.38$ & \multicolumn{1}{|c}{$17.78$} & \multicolumn{1}{|c}{$1.88$}
& \multicolumn{1}{|c}{$1.82$} & \multicolumn{1}{|c|}{$1.87$} \\ \hline
$375$ & $29.44$ & \multicolumn{1}{|c}{$16.68$} & \multicolumn{1}{|c}{$-0.71$}
& \multicolumn{1}{|c}{$-0.69$} & \multicolumn{1}{|c|}{$-0.71$} \\ \hline
$400$ & $28.58$ & \multicolumn{1}{|c}{$15.69$} & \multicolumn{1}{|c}{$-3.18$}
& \multicolumn{1}{|c}{$-3.07$} & \multicolumn{1}{|c|}{$-3.20$} \\ \hline
$425$ & $27.79$ & \multicolumn{1}{|c}{$14.79$} & \multicolumn{1}{|c}{$-5.54$}
& \multicolumn{1}{|c}{$-5.34$} & \multicolumn{1}{|c|}{$-5.61$} \\ \hline
$450$ & $27.06$ & \multicolumn{1}{|c}{$13.99$} & \multicolumn{1}{|c}{$-7.82$}
& \multicolumn{1}{|c}{$-7.51$} & \multicolumn{1}{|c|}{$-7.94$} \\ \hline
$475$ & $26.39$ & \multicolumn{1}{|c}{$13.25$} & \multicolumn{1}{|c}{$-10.02$%
} & \multicolumn{1}{|c}{$-9.59$} & \multicolumn{1}{|c|}{$-10.21$} \\ \hline
$500$ & $25.77$ & \multicolumn{1}{|c}{$12.58$} & \multicolumn{1}{|c}{$-12.14$%
} & \multicolumn{1}{|c}{$-11.60$} & \multicolumn{1}{|c|}{$-12.41$} \\ \hline
\end{tabular}
\end{center}

\newpage \noindent \textbf{Table 2.} Calculated values of the pole
parameters $D$ and $T_{0}$, and of the accuracy of the phase-shift
description $\Delta $ for different energy intervals $0\leqslant T\leqslant
T_{\max }$.

\begin{center}
\begin{tabular}{|c|c|c|c|}
\hline
$T_{\max }\,,\,$ & $D\,,$ & $T_{0}\,,\,$ & $\Delta \,,$ \\ 
$\,$MeV & fm$^{2}$ & MeV & deg \\ \hline
$10$ & $-0.2007$ & $413.5$ & $0.0056$ \\ \hline
$20$ & $-0.2147$ & $386.6$ & $0.0057$ \\ \hline
$30$ & $-0.2172$ & $382.1$ & $0.0058$ \\ \hline
$40$ & $-0.2236$ & $371.2$ & $0.0091$ \\ \hline
$50$ & $-0.2303$ & $360.4$ & $0.014$ \\ \hline
$60$ & $-0.2355$ & $352.4$ & $0.020$ \\ \hline
$80$ & $-0.2416$ & $343.5$ & $0.029$ \\ \hline
$100$ & $-0.2432$ & $341.3$ & $0.032$ \\ \hline
$150$ & $-0.2412$ & $344.1$ & $0.061$ \\ \hline
$200$ & $-0.2356$ & $352.3$ & $0.16$ \\ \hline
$250$ & $-0.2315$ & $358.5$ & $0.26$ \\ \hline
$300$ & $-0.2290$ & $362.4$ & $0.33$ \\ \hline
$350$ & $-0.2279$ & $364.2$ & $0.36$ \\ \hline
$368$ & $-0.2256$ & $368.0$ & $0.44$ \\ \hline
\end{tabular}
\\[0pt]
\end{center}

\newpage \noindent \textbf{Table 3.} Triplet low-energy scattering
parameters obtained by using the data of the partial-wave analysis performed
by the GWU group (versions GWU P1 and GWU P2 correspond to the pole
approximation at the shape-parameter values of $v_{2}=0.163\,$fm$^{3}$ and $%
v_{2}=0.168\,$fm$^{3}$, respectively) and the data of the partial-wave
analysis performed by the Nijmegen group \lbrack 50, 57\rbrack\ (Nijm
version).

\begin{center}
\begin{tabular}{|c|c|c|c|c|c|c|}
\hline
Version & $a_{t}$, fm & $r_{t}$, fm & $v_{2}$, fm$^{3}$ & $v_{3}$, fm$^{5}$
& $v_{4}$, fm$^{7}$ & $v_{5}$, fm$^{9}$ \\ \hline
GWU P1 & $5.4030$ & $1.7494$ & $0.163$ & $0.037$ & $0.0083$ & $0.0019$ \\ 
\hline
GWU P2 & $5.4030$ & $1.7494$ & $0.168$ & $0.038$ & $0.0085$ & $0.0019$ \\ 
\hline
Nijm & $5.420$ & $1.753$ & $0.040$ & $0.672$ & $-3.96$ & $27.1$ \\ \hline
\end{tabular}
\end{center}

\newpage \noindent \textbf{Table 4.} Triplet dimensionless shape parameters $%
p_{n}$ obtained by using the data of the partial-wave analysis performed by
the GWU group (version GWU corresponds to the pole approximation P1 at the
shape-parameter value of $v_{2}=0.163\,$fm$^{3}$ ) and the data of the
partial-wave analysis performed by the Nijmegen group \lbrack 50, 57\rbrack\
(Nijm version).

\begin{center}
\begin{tabular}{|c|c|c|c|c|}
\hline
Version & $p_{2}$ & $p_{3}$ & $p_{4}$ & $p_{5}$ \\ \hline
GWU & $0.0304$ & $0.0023$ & $0.00017$ & $0.000012$ \\ \hline
Nijm & $0.00743$ & $0.0406$ & $-0.0778$ & $0.173$ \\ \hline
\end{tabular}
\end{center}

\newpage \noindent \textbf{Table 5.} Features of the deuteron according to
calculations within various approximations of the effective-range expansion
(ER, SP, $v_{n}$) and in the pole approximation P1.

\begin{center}
\begin{tabular}{|c|c|c|c|c|}
\hline
Version & $\varepsilon _{d}$, MeV & $\rho _{d}$, fm & $C_{d}$ & $A_{S}$, fm$%
^{-1/2}$ \\ \hline
ER & $2.2387$ & $1.7494$ & $1.2979$ & $0.8846$ \\ \hline
SP & $2.2236$ & $1.7145$ & $1.2876$ & $0.8761$ \\ \hline
$v_{3}$ & $2.2237$ & $1.7151$ & $1.2878$ & $0.8763$ \\ \hline
$v_{4}$ & $2.2237$ & $1.7151$ & $1.2878$ & $0.8763$ \\ \hline
$v_{5}$ & $2.2237$ & $1.7151$ & $1.2878$ & $0.8763$ \\ \hline
P1 & $2.2237$ & $1.7151$ & $1.2878$ & $0.8763$ \\ \hline
Experiment & $2.224589$ & $1.7251$ & $1.2904$ & $0.8781$ \\ \hline
\end{tabular}
\end{center}

\newpage

\vspace*{2cm}

\begin{center}
\unitlength=0.24pt 
\begin{picture}(1317,953)
\put(0,953){\special{em:graph 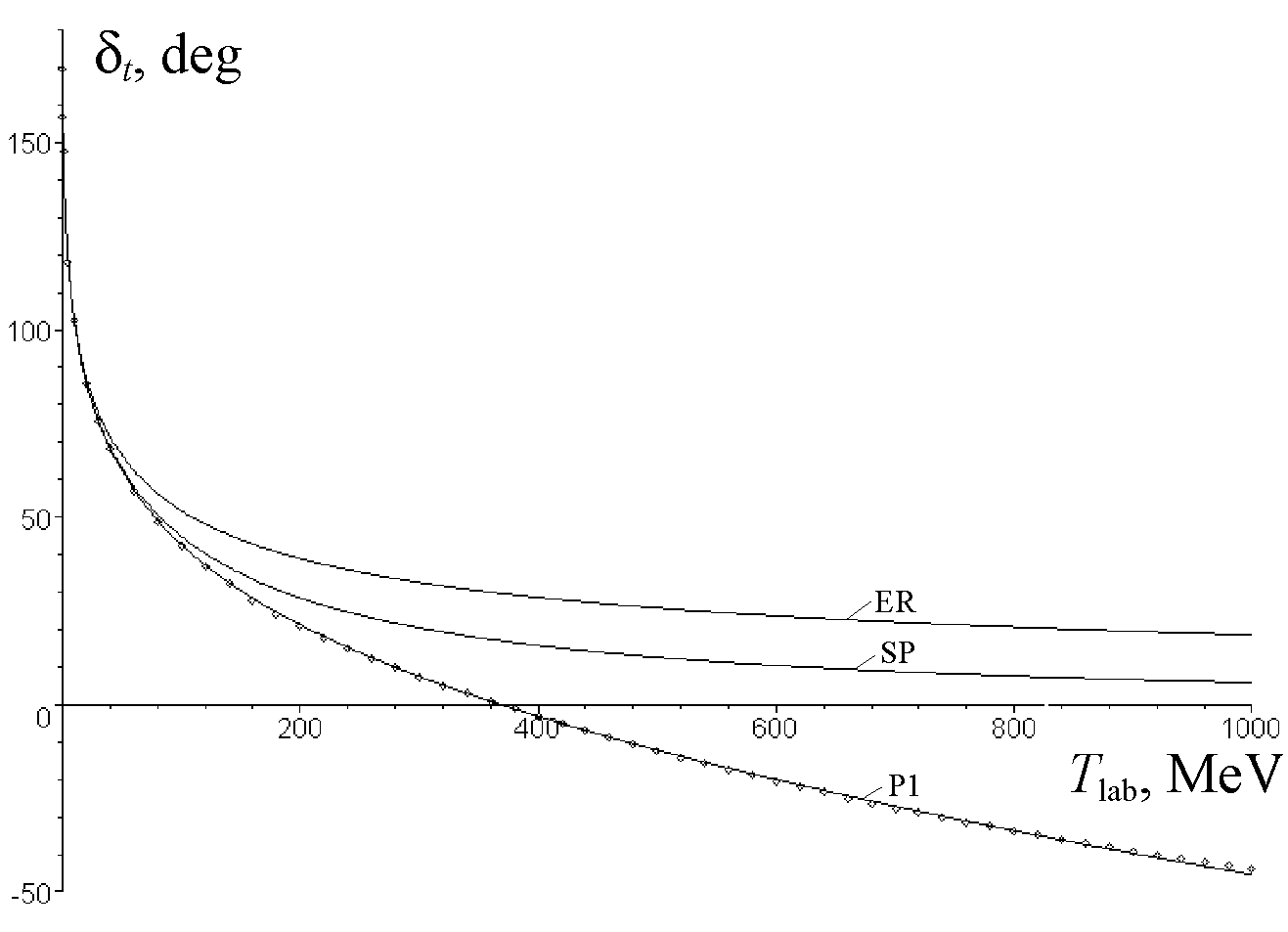}}
\end{picture}
\end{center}

\noindent \textbf{Fig. 1. }Triplet phase shift for neutron-proton scattering
as a function of the laboratory energy according to calculations within the
effective-range (ER), the shape-parameter (SP), and the pole (P1)
approximation. The points represent experimental data borrowed from \lbrack
39\rbrack . \ 

\newpage

\vspace*{2cm}

\begin{center}
\unitlength=0.24pt 
\begin{picture}(954,953)
\put(0,953){\special{em:graph 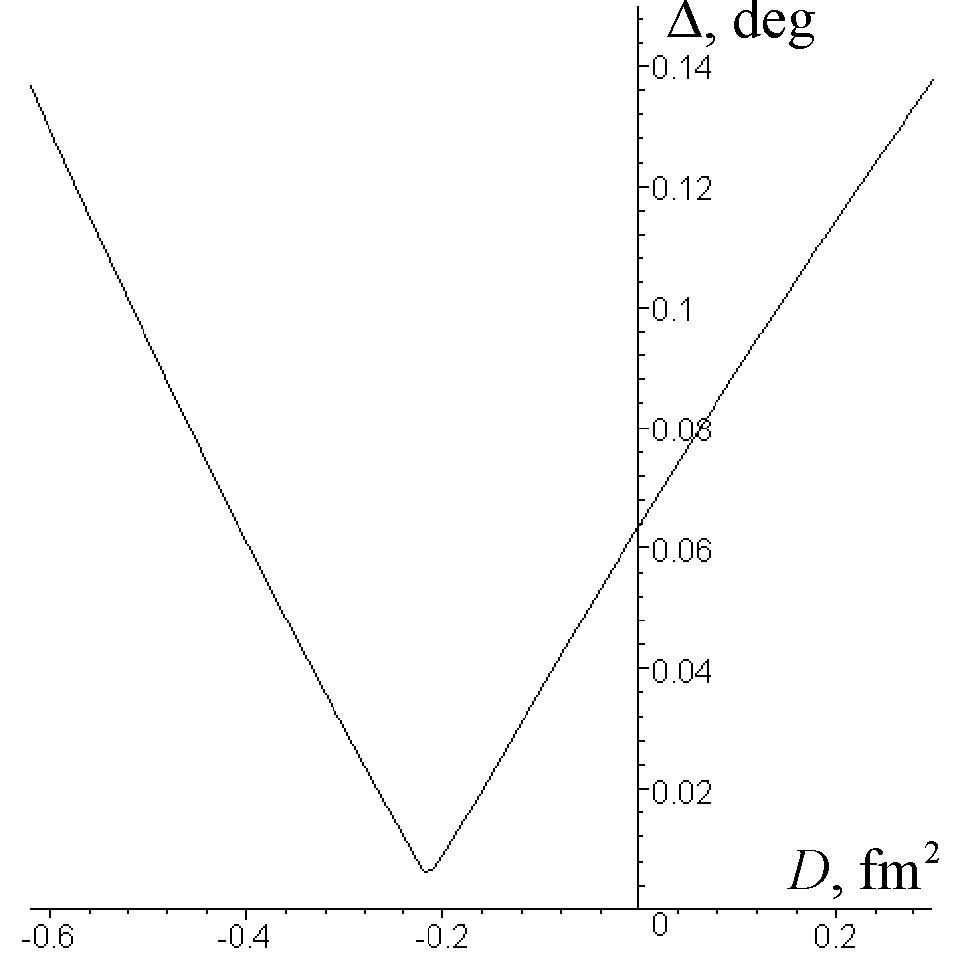}}
\end{picture}
\end{center}

\noindent \textbf{Fig. 2. }Dependence of the accuracy of the phase-shift
description $\Delta $ on the pole parameter $D$ for the energy interval $%
0\leqslant T\leqslant 20\,$MeV$\,$.

\newpage

\vspace*{2cm}

\begin{center}
\unitlength=0.24pt 
\begin{picture}(1239,953)
\put(0,953){\special{em:graph 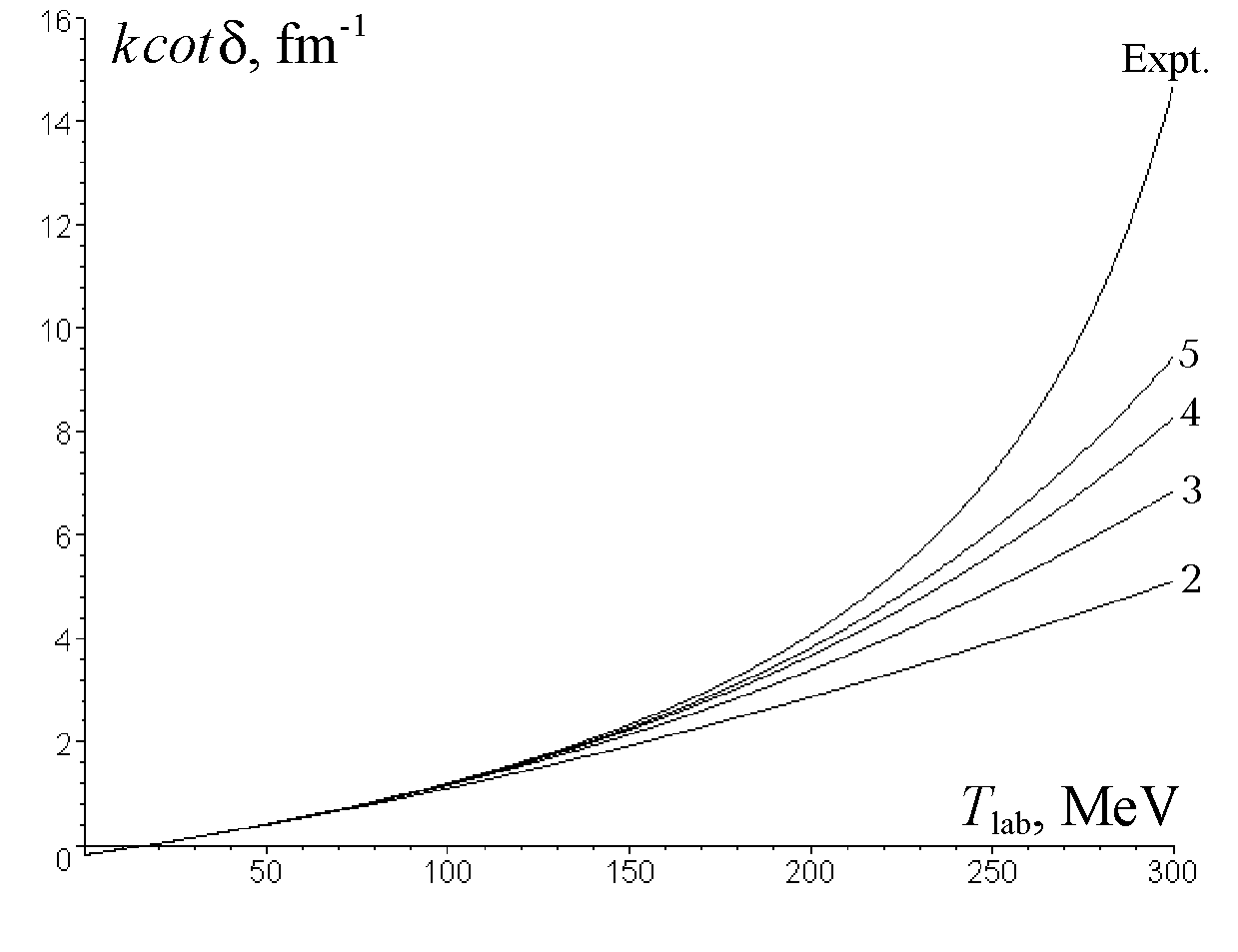}}
\end{picture}
\end{center}

\noindent \textbf{Fig. 3. }Triplet effective-range function $k\cot \delta $
versus the laboratory energy for various polynomial approximations of the
effective-range expansion (11). The figures on the curves indicate the order
of approximation. The upper curve corresponds to experimental data borrowed
from \lbrack 39\rbrack .

\end{document}